# Methods for scoring the collective effect of SNPs: Minor alleles of common SNPs quantitatively affect traits/diseases and are under both positive and negative selection

Dejian Yuan[1#], Zuobin Zhu[1#], Xiaohua Tan[1], Jie Liang[1], Ceng Zeng[1], Jiegen Zhang[2], Jun Chen[2], Long Ma[1], Ayca Dogan[3], Gudrun Brockmann[3], Oliver Goldmann[4], Eva Medina[4], Amanda D. Rice[5], Richard W. Moyer[5], Xian Man[1], Ke Yi[1], Yanke Li[1], Qing Lu[1], Yimin Huang[1], Dapeng Wang[6], Jun Yu[6], Hui Guo[1], Kun Xia[1], and Shi Huang[1*]

[1]State Key Laboratory of Medical Genetics, Central South University, 110 Xiangya Road, Changsha, Hunan, China 410078

[2]High Performance Computing Center, Modern Educational Technology Center, New Campus, Central South University, Changsha, Hunan 410083, China

[3]Department of Crop and Animal Sciences, Faculty of Agriculture and Horticulture, Humboldt-Universität zu Berlin, Invalidenstraße 42, 10115 Berlin, Germany

[4]Infection Immunology Group, HZI – Helmholtz Centre for Infection Research, Inhoffenstraße 7, 38124 Braunschweig, Germany

[5]Department of Molecular Genetics and Microbiology, University of Florida, 1600 SW Archer Rd., Gainesville, FL 32610, USA

[6]CAS Key Laboratory of Genome Sciences and Information, Beijing Institute of Genomics, Chinese Academy of Sciences, Beijing 100029, China

*Correspondence to: Shi Huang, huangshi@sklmg.edu.cn

#These authors contributed equally to this work.

**Abstract**:

Most common SNPs are popularly assumed to be neutral. We here developed novel methods to examine in animal models and humans whether extreme amount of minor alleles (MAs) carried by an individual may represent extreme trait values and common diseases. We analyzed panels of genetic reference populations and identified the MAs in each panel and the MA content (MAC) that each strain carried. We also analyzed 21 published GWAS datasets of human diseases and identified the MAC of each case or control. MAC was nearly linearly linked to quantitative variations in numerous traits in model organisms, including life span, tumor susceptibility, learning and memory, sensitivity to alcohol and anti-psychotic drugs, and two correlated traits poor reproductive fitness and strong immunity. Similarly, in Europeans or European Americans, enrichment of MAs of fast but not slow evolutionary rate was linked to autoimmune and numerous other diseases, including type 2 diabetes, Parkinson's disease, psychiatric disorders, alcohol and cocaine addictions, cancer, and less life span. Therefore, both high and low MAC correlated with extreme values in many traits, indicating stabilizing selection on most MAs. The methods here are broadly applicable and may help solve the missing heritability problem in complex traits/diseases.

*Abbreviations*:

MAF: minor allele frequency

MAC: minor allele content.

**Introduction:**

Although past studies of complex traits/diseases have met with great success in identifying a number of significant genetic variants in various traits and diseases, such variants usually account for only a small fraction of the total trait variation and their functional roles typically remain unclear, which has led to the missing heritability problem (Burton, et al., 2007; Conrad, et al., 2010; Manolio, et al., 2009; O'Donovan, et al., 2008; Purcell, et al., 2009; Shi, et al., 2009; Stefansson, et al., 2009; Teslovich, et al., 2010). The focus on searching for a few major effect variants is under the null hypothesis in the field of population genetics that the majority of genetic variations are neutral and hence irrelevant to phenotypes. Notably, however, while the assumption of neutrality for most variations has often passed tests by sequence-alignment based informatics approaches, such methods usually have their own set of uncertain assumptions, including assuming certain DNAs to be neutral such as synonymous sites and transposon-element derived sequences, and are therefore not truly conclusive tests free of neutral or uncertain assumptions (Fay, et al., 2002; Ponting and Hardison, 2011). In contrast, direct tests by experimental science suggest at least 80% functional human genome (Dunham, et al., 2012).

While too little genetic variations are known to hurt adaptive capacities, it is much less appreciated whether too much may exceed an organism's maximum level of tolerable entropy, given that mutations are after all random and disorderly in origin. We here developed two novel methods to test whether the total amount of SNPs carried by an individual are under stabilizing selection. First, we made use of multiple panels of genetic reference populations or recombinant inbred lines (RILs) that provide a powerful means to study the genetic basis of complex phenotypes (Ayroles, et al., 2009; Brem, et al., 2002; Perlstein, et al., 2007; Philip, et al., 2010; Philip, et al., 2011; Seidel, et al., 2008; Taylor, 1978; Vinuela, et al., 2010). The RIL panels are derived from breeding of parental strains differing in phenotypes and genotypes. The F1 and F2 or up to F10 progenies are intercrossed to maximize random recombination and hence allelic diversity in the offspring, which were then randomly selected for inbreeding up to 20 generations to generate the final panel of RILs homozygous for almost all variants or SNPs. During the random mating and subsequent inbreeding process, there are ample opportunities for neutral variants to drift and for non-neutral variants to be positively and negatively selected. Immunity against pathogens is essential for survival and depends on allelic diversity, which would positively select for enrichment of variants. On the other hand, individuals may die or be aborted before birth due to abnormal development caused by enrichment of deleterious variants. While the population size of a RIL panel is small, which favors neutral drift, the actual size of the offspring population of the original parents is much greater and includes many that died because of poor immunity or were never born due to *en utero* negative selection.

If a trait is determined by multiple loci and robust to minor perturbations, one may expect that the trait may be genetically affected in two mutually non-exclusive ways. One is a major effect mutation in one of these loci that alters a component of a multi-component pathway. Alternatively, it may take a large amount of minor effect mutations in multiple loci to harm the trait, while such mutations individually or even a small collection of them may have few discernable effects or even beneficial effects. Furthermore, the variation in the number of such minor effect mutations may account for any quantitative variation in a trait, which characterizes most complex traits. For example, the more the variants the better the immunity upto a point when too much variants may start to hurt other traits or be cancer prone.

For any given panel of RILs, most SNPs would show MAs that are carried by less than half of the strains in the panel and the strains would differ in the contents of MAs that each carries. We defined "MA contents (MAC)" as the number of MAs in an individual divided by the number of SNPs scanned. Different from MAF, MAC is an individual measure. One predicts that strains with higher MAC should be similar to those with lower if the neutrality assumption is true. Alternatively, they should show poorer measurements in certain adaptive traits while better in certain others that are known to depend on allelic diversity if the natural selection model is true. We here tested these predictions by performing new trait analysis experiments as well as by using the large collection of data accumulated in the past several decades for genetic reference populations.

Taking advantage of published SNP datasets from genome wide association studies (GWAS) of numerous human diseases that were typically genotyped at ~450-900K common SNPs, our second approach was to determine whether the MAC of a patient is different from that of a control. We focused on diseases for which results in model organisms can provide independent evidence.

## Results:

### MA distribution profiles in genetic reference populations

We calculated MAF for each scanned SNP in a panel of genetic reference population. We then calculated MAC for each strain of a panel and plotted the MAC distribution curve (Figure 1 and Supp. Table S1 for strain descriptions). Great variations in MAC (~0.2 to ~0.7) were observed for *C. elegans* RILs from either the Kruglyak or the Kammenga laboratory (Figure 1A and B) (Seidel, et al., 2008; Vinuela, et al., 2010), the yeast segregant panel (Figure 1C) (Brem, et al., 2002; Perlstein, et al., 2007), and the BXD mouse RIL panel (Figure 1F) (Philip, et al., 2010; Taylor, 1978). Relative to these RILs, *D. melanogaster* inbred panel derived from the wild showed lower MAC and smaller variation range (Figure 1D) (Ayroles, et al., 2009). RILs that were only partially inbred such as the Collaborative Cross (CC) G2F7 mouse panel that has been inbred for only 7 generations also showed small variation range in MAC (Figure 1E) (Philip, et al., 2011). For certain panels with large variations, an abrupt turn at the ends of the distribution curve, especially the higher end, was apparent, indicating an under-representation and hence lower survival success of strains with low or high MAC (Figure 1A-D, F). The population distribution of MAC showed a bell curve as expected (Supp. Figure S1). For calculating MAC, the number of informative SNPs used for each panel ranged from ~0.12K to ~151K. Since the SNPs are a random sampling of the genome, the number of SNPs used should not significantly affect the calculation of MAC. Indeed as shown for the BXD mouse panel, MAC calculated from ~51K SNPs were highly similar to those calculated using two different non-overlapping sets of 1K SNPs randomly selected from the ~51K (Supp. Figure S2).

**MAC correlates with quantitative variations in complex traits in model organisms**

To determine whether MAC may affect reproductive fitness, we examined brood size of 42 *C. elegans* RILs from the Kruglyak laboratory with Hawaii (HW) *npr-1* genotype and 62 RILs with N2 *npr-1* genotype (Figure 2A-B). Their parental strains Hawaii CB4856 and Bristol N2 differ in *npr-1* by one major effect SNP (F215V). Higher MAC was linked with lower brood size in a nearly linear fashion, with its effect stronger in HW *npr-1* background (Figure 2A-B). The deleterious effect of higher MAC on reproductive traits was confirmed in three other RIL panels in mouse and rat, BXD, CC(G2F6), and BXHHXB (Supp. Table S2). In addition, higher MAC was linked with lower life span in *C. elegans* and mouse (Supp. Table S2), less startle response in *D. melanogaster* (for males, Spearman r = - 0.23, $p = 0.004$) (Ayroles, et al., 2009), and more chill coma response in *D. melanogaster* (for females, Spearman r = 0.22, $p = 0.007$) (Ayroles, et al., 2009).

There are 3664 traits for the BXD mouse panel of 89 strains characterized by numerous studies (some unpublished) with data archived at GeneNetwork (Taylor, et al., 1999; Wang, et al., 2003). For $p < 0.0001$ and 0% FDR as determined by 100 permutations, there were 15 traits linked with MAs by Pearson and 15 by Spearman analysis, and there were 297 traits with $P<0.05$ and 60% FDR by Spearman analysis (Supp. Table S3). A few examples of significant results include lower maximum threshold to ethanol induced ataxia (Figure 3A) and lower blood ethanol concentration in males 20 min after ethanol injection (Figure 3B). A number of related neurological traits were linked with higher MAC, including smaller methamphetamine-induced body temperature change, slower reversal learning, higher sensitivity to pain, less open field rearing behavior, more anxiety as assayed by the light-dark box method, less anxiety as assayed by the elevated plus maze assay, more depression as assayed by duration of immobility in a tail suspension test, more saccharin preference, and some cocaine related traits (Supp. Table S3 and S4). However, MAC was not linked with morphine-induced difference in locomotion. Other traits more definitively linked with higher MAC ($p < 0.0001$ and 0 FDR) include higher adrenal weight, higher deoxycorticosterone level in cerebral cortex, and less value in the unpublished traits 274 and 372 of the AUW-BXD series that were significantly correlated with slower reversal learning (Supp. Table S3).

Most of the 3664 traits in Supp. Table S3 were scored for less than half of the panel and different sets of strains were often used for scoring different traits. After filtering out traits and strains with too many missing data, we were able to perform multivariate regression analysis on 9 traits, which identified 3 significant associations, including blood ethanol concentration, adrenal zona fasciculata width, and hair coat color (Supp. Table S5).

A number of traits were repeatedly found linked with MAC in different panels of RILs (Supp. Table S2). One was tumor susceptibility (Figure 4 and Supp. Table S2). The effect of MAC in urethane induced lung tumor was only apparent when *kras2* oncogene was wild type (Figure 4A vs 4B)(Ryan, et al., 1987). There were also traits such as blood pressure that were repeatedly not found associated with MAC (Supp. Table S2).

MAC also consistently associated with traits linked to obesity, type 2 diabetes (T2D) and Parkinson's disease (PD). In BXHHXB rat, more MAs were linked with higher glucose level after high fructose diet (Figure 5A) and lower serum dopamine level (Figure 5B). In high fat diet fed BXD mice, higher MAC correlated with traits associated with obesity and T2D, including higher resistin level and more body weight increase, increased food intake, higher oxygen consumption, and lower transferring saturation (Supp. Table S3 and S4).

The expectation of a link between immunity and allelic diversity was confirmed in two panels of mouse RILs (BXD and BXH) with higher MAC associated with stronger immunity (Table 1).

For the yeast segregant panel, we analyzed the published 316 response profiles to 92 drugs and chemical compounds including 18 Food and Drug Administration (FDA)-approved drugs (Perlstein, et al., 2007). By SAM analysis, we identified 12 traits at 0 FDR among 316 drug response traits. All 12 traits showed more growth inhibition in strains with higher MAC, suggesting less reproductive abilities of these segregants under certain environmental conditions but never more under any of the 316 conditions examined here. Seven among these involved 4 drugs that are FDA-approved antipsychotic and antidepression drugs (sertraline, trimeprazine, chlorpromazine, and trifluoperazine), and one involved the FDA-approved breast cancer drug Tamoxifen (Supp. Table S6). The remaining four traits involved three chemical compounds and there was an enrichment of FDA-approved drugs among chemical compounds with effects on strains with higher MAC (5 of 18 vs 3 of 74, $p < 0.01$, chi-square test).

**Trait specific set of MAs**

While many different traits were linked with MAC, most of them did not share the same set of MAs since a trait was only correlated with some but not most other traits (Supp. Table S4). For example, blood ethanol concentration (BEC) strongly correlated with resistin level but not pain response and open field rearing behavior (Supp. Table S4). To confirm among the MAC-linked traits that correlated traits share more MAs than non-correlated traits, we developed an approach to identify trait specific set of MAs as described in the Methods. From 51469 SNPs originally used for calculating MAC for the BXD panel, we identified a BEC-specific set of 30336 SNPs. When the MAC value was calculated using the BEC specific set, the BEC trait was strongly linked with MAC (Spearman r = - 0.66, $p = 0.0004$) and so was its related trait resistin level (r = 0.53, $p = 0.008$). No strong association was noted for its two non-related traits, rearing behavior (r = - 0.4, $p = 0.06$) and pain (r = - 0.46, $p = 0.03$), thus demonstrating the specificity of the BEC-specific set of SNPs. In contrast, using the non-trait specific set of 51469 SNPs, two non-related traits, resistin level and pain response, scored the best correlation with MAC.

We next examined whether poor reproductive fitness and strong immunity are correlated traits. The MAC-linked reproductive trait in BXD mice is uterus horn length at maturity as shown in Supp. Table S2. There was a correlation between this trait and formation of secondary dermal lesions upon Ectromelia virus infection of footpad (Pearson r = 0.40, $p = 0.028$ for mixed sexes). Such correlation suggests sharing of SNPs in reproduction and immunity.

**MAC in common diseases**

The above results in model organisms suggest that higher MAC may contribute to certain human diseases. We made use of published GWAS datasets of Europeans or European Americans (EA), which generally have cases well matched with controls in terms of population straitification. We identified the MAs of informative autosomal SNPs in the control population of each study and determined the average distance of either cases or controls to the set of MAs. Smaller distance to the MA set means greater MAC. For most of the diseases here, we studied at least two independent datasets in order to verify a positive link.

As shown in Table 2, all diseases studied except hypertension and coronary artery disease (the former is a high risk factor for the latter) consistently showed higher MAC in cases (most with $p < 0.01$). Higher MAC correlated more often with higher heterozygosity and greater pairwise genetic distance, as may be expected (Table 2). While we only had one dataset for each of the three autoimmune diseases studied here, type 1 diabetes, inflammatory bowel disease, and rheumatoid arthritis, the fact that all three showed linkage with higher MAC indicates a consistent pattern for autoimmune disease as a whole. In all diseases with higher MAC in cases, there was at least one cohort showing significant Spearman correlation between higher MAC and cases ($r > 0.036$, $p < 0.05$), where we mixed cases and controls with case status as 1 and control as 0 and tested how well the MAC of each subject correlated with case status. The strongest correlation was found in one PD cohort (Spearman r = 0.21, $p < 0.0001$) and the weakest for rheumatoid arthritis (Table 2). We then picked the Wellcome Trust Case Control Consortium (WTCCC) T2D dataset as an example to see how the MAC linkage with case status compares to top SNP hits previously identified by GWAS that typically have trend $p < 5 \times 10^{-7}$ (Burton, et al., 2007). The Spearman correlation between MAC and case status in WTCCC T2D ($r = 0.085$, $p < 0.0001$) was similar to that between genotype and case status for the top T2D-associated SNP rs4506565 with reported trend $p = 5.68 \times 10^{-13}$ ($r = 0.105$, $p < 0.0001$) and rs9939609 with trend $p = 5.24 \times 10^{-8}$ ($r = 0.076$, $p = 0.0001$) (Burton, et al., 2007). As a further control, two randomly chosen control populations from PD data showed no significant difference in MAC.

Although the GWAS datasets have controlled for population stratification, we used a novel strategy to confirm it. Slow evolving sequences are more likely to follow the infinite sites model of the neutral theory while fast evolving ones are likely to have reached saturation where new mutations would occur at old sites that have encountered mutations before. Therefore, one should observe a dramatic difference between SNPs of fast and slow evolutionary rates if indeed time has been long enough for most fast evolving SNPs to reach saturation. As shown in Table 2, non-synonymous SNPs located in the slowest evolving proteins showed higher MAC in controls relative to cases in 12 of 21 datasets with two being significant (dbGaP T2D female and male cohorts, Table 2). The other 9 datasets showed higher MAC in cases also with two being significant (dbGaP major depression cohort 2 and lung cancer H610). The nearly even split of the datasets into those with higher MAC in controls versus those in cases was to be expected if the cases and controls were well matched. Thus, the higher MAC in 19 of the 21 sets of cases as measured by most common SNPs, which mostly locate in relatively fast evolving sequences, is not due to population stratification (difference between slow and fast SNPs was significant: 19 of 21 vs 9 of 21, $p < 0.01$, chi-square test). Also, if MAs in fast evolving sequences have no roles in human diseases in general, we should not expect a significant excess of datasets with higher

MAC in cases (19 of 21, $p < 0.01$, chi-square test). Indeed, results with slow SNPs showed the expected equal distribution of MAs in cases and controls (9 datasets with higher MAC in cases vs 11 with higher MAC in controls, $p > 0.05$, chi-square test).

This difference between slow and fast SNPs was not due to the trivial reason that the number of slow SNPs analyzed was much smaller than that of fast SNPs, because randomly selected sets of small number of fast SNPs gave similar results as the large sets of fast SNPs. For example, for PD cohort1, three randomly selected non-overlapping sets of 168 fast SNPs all showed significantly more MAs in cases with $p < 2.0E-4$, whereas in contrast the same number of slow SNPs showed insignificantly higher MAC in controls as shown in Table 2.

Finally, we examined MAC relationship with clinical traits in cases and controls where such information was available. Among alcohol addiction cases, MAC correlated with 7 traits of cocaine addiction but not with marijuana and opiate addictions (Supp. Table S7). In both cases and controls in the alcohol addiction study, higher MAC was linked with lower education grade level achieved, consistent with its role in slower learning and memory as above found for mice. The PD dataset has age of death information for some of the control subjects and revealed a link between more MAC and shorter life span. The PD control dataset also showed a link between more MAC and higher intake of over the counter non-steroid anti-inflammatory drugs (NSAIDs), confirming the link between MAC and over-active immune system as above observed for autoimmune diseases.

**Discussion:**

Our results show a connection between MA contents and numerous traits and diseases. MAC was linked with poorer rather than better performance in many adaptive traits. Lower reproductive fitness such as smaller uterus may be sufficient to explain the lower frequency of some of these MAs. Negative selection *en utero* may also do so, and the deleterious effects of these MAs on some adult traits/diseases may reflect pleiotropy. Thus, the minor nature of MAs linked with late onset diseases such as T2D and PD may reflect negative selection *en utero* rather than by these diseases per se. In contrast, the link between higher MAC and better immunity and the inverse correlation between immunity and reproduction indicate simultaneous positive selection of the negatively selected MAs and explain why a common MA should be common rather than rare.

Immunity and reproduction may not be the only traits to make possible both positive and negative selection of an MA. A more general way is the negative selection of extreme MAC associated with extreme measurements of numerous quantitative traits (Figure 6). The adaptive advantages of allelic diversity are well appreciated by past studies, including conferring a wide range of quantitative variations in nearly all complex traits. For most quantitative traits, the two extremes of trait measurements, either high or low, represent suboptimum population minorities, and are negatively selected by way of ultimately affecting reproductive fitness as most traits could be somehow related to reproduction given its central role in evolution. The two extremes would be represented by high and low MAC if the quantitative variations in such traits are related to the additive effect of MAs as shown here for numerous traits as well as by the observation that traits known to have large number of additive QTLs are more likely to be affected by MAC (Zhu, et al., 2013). Thus both high and low MAC would be negatively selected, which means both positive and negative selection for these MAs. For example, too weak or too strong immunity would be both negatively selected. The same goes for too low or too high anxiety (or depression, adrenal weight, deoxycorticosterone level, etc). However, the extreme value of a trait variation linked with low MAC should be less harmful than the other extreme linked with high MAC since after all these alleles are minor or on the whole slightly deleterious. Thus, while diseases linked with low MAC are expected to exist (congenital immunodeficiency syndromes may be one), they may be less common.

This study here analyzed more than 4000 traits for associations with MAC in animal models and humans. Hundreds of associations passed the significance value of $P < 0.05$. Most of these should be considered as results of an exploratory study needing future verification. Using a multiplicity adjustment method such as the Bonferronni correction here would eliminate most of these associations as insignificant false positives but the correction is widely known to have a few fallacies to deem its use here improper (Perneger, 1998) In addition to defying common sense, the Bonferronni correction increases type 2 errors or the chance of false negatives especially when involving extremely large number of tests as was the case here. Typical GWAS studies do make multiple test corrections based on uncertain assumptions and cannot draw meaningful conclusions on SNPs with P values between 0.05 and $5 \times 10^{-7}$ (Burton, et al., 2007). But as shown here by the MAC concept/methods, most such SNPs may in fact be significant. Most animal experiments have small sample sizes due to practical and financial reasons. The value of our study is to give the community a select list of MAC-linked traits worthy of future confirmatory studies. A future study testing one of the linked traits here would only have one hypothesis to test and thus only need to achieve standard P value ( $<0.05$).

Do the results here mean an additive effect of large numbers of MAs in MAC action and hence non-neutrality of most MAs? Many major effect risk alleles of diseases are known to be minor alleles (Park, et al., 2011), which may plausibly imply that the effect of MAC may be mediated by a few known major effect risk alleles rather than large numbers of minor effect MAs.

But this may not be the case. The effect of MAC was in fact abolished or weakened by major effect MAs such as *kras2* mutation in lung cancer or *npr-1* mutation in brood size. Furthermore, MAC preferentially affects traits with larger number of known additive QTLs (Zhu, et al., 2013). It is self-evident that the more the number of QTLs involved in a trait, the less the individual effect of each QTL on the trait. These results also mean that the effect of MAC would become much more obvious and significant after filtering out individuals with major effect MAs such as *kras2* or *npr-1* mutation, which is exactly the opposite of what one would predict from the notion of MAC involving known major effect MAs. Thus, MAC-linked diseases are expected to have more additive minor effect SNPs as risk alleles than those not linked to MAC. The individual effect of such SNPs may not be possible for existing methods like GWAS to detect. In this sense, the concept and methods of MAC here may help solve the "missing heritability" problem of some complex diseases/traits. Major effect risk alleles and MAC may each account for different fractions of a case population. The size of the fraction accounted by MAC needs to be determined by future studies using trait/disease specific set of MAs and filtering out cases linked with major effect MAs. We predict further that cases associated with major effect risk alleles should have different defects and hence be treated differently from those linked with MAC.

The results here in model organisms and humans suggest that most common MAs in any species are nearly neutral or slightly deleterious resulting from both positive and negative selection. As new mutations constantly introduce more MAs, a slightly more dominant negative selection relative to positive selection is absolutely necessary in order to prevent allelic diversity to be over an optimum limit. In contrast, positive selection may not always be necessary for MAs to reach an optimum amount as mutations plus time or neutral drift can do so too if only slowly. At the optimum level of allelic diversity, the overall slightly deleterious nature of MAs would be in homeostasis with the slightly beneficial nature of major alleles. The optimum concept here means a Pareto optimum or simply the best that can be achieved due to a balance between positive and negative selection at a particular time point under a specific level of epigenetic or organismal complexity (Shoval, et al., 2012). As time and complexity changes, the optimum level of nucleotide diversity will also change. While what determines genetic diversity has been a long-standing unsolved puzzle (Leffler, et al., 2012), the results here suggest a critical role of the internal system construction requirements of the species or its level of complexity.

The results here that cases have more MAC or nucleotide diversity provides further direct evidence for the inverse relationship between genetic diversity and epigenetic complexity and the Maximum Genetic Diversity (MGD) hypothesis, a more complete account of evolutionary and hereditary phenomena that views most random mutations as entropy generating rather than neutral (Hu, et al., 2013; Huang, 2008; Huang, 2009; Huang, 2010; Huang, 2012). The MGD absorbs all the proven virtues of the neutral theory and natural selection in describing the linear phase of evolution where genetic distance is linearly related to time, but considers most sequence divergence observed today are in fact at optimum or maximum levels. Certain key evidence of maximum divergence, such as the genetic equidistance result of Margoliash, has been unfortunately misinterpretated by the neutral perspective for nearly half of a century (Hu, et al., 2013; Huang, 2008; Huang, 2009; Huang, 2010; Huang, 2012; Margoliash, 1963). A useful standard for judging the validity of a theory about the past may be to see how relevant it is to solving problems of today, since knowing the past should help know the present. The studies here were inspired by the MGD hypothesis and illustrate its practical value to contemporary problems. In contrast, by assuming ~90% of the human genome as junks or neutral (Ponting and Hardison, 2011), the neutral theory has self rendered itself irrelevant to the information/traits associated with such DNAs and not unexpectedly cannot help with, if not in fact creating, the missing heritability problem.

Ohta's theory of slightly deleterious or nearly neutral mutations works nearly as well as Kimura's neutral theory in explaining the behavior of genetic polymorphisms (Ohta, 1973). But, a serious problem with it from the perspective of the existing framework as pointed out by Nei and Kumar (Nei and Kumar, 2000) is that "if slightly deleterious mutations are abundant and are fixed in the populations continuously, the genes will gradually deteriorate and eventually lose their original function." The results here show that these mutations cannot accumulate without a limit and are negatively selected in bulk, as predicted by the MGD. However, the nearly neutral theory predicts no such limit on genetic diversity, and is hence disproven by the results here.

Similar to the laws in hard sciences such as physics and mathematics for which the foundations are intuitions or axioms, both Darwin's natural selection and Kimura's neutral theory, as applied in population genetics for sequences yet to reach optimum diversity, are also self-evidently true or could be viewed as a deduction of an intuition in construction that random errors/noises within tolerable levels can be neutral, deleterious, or beneficial depending on circumstances. Limited genetic diversity at optimum level are more likely to be beneficial than deleterious because they are within tolerable levels and confer strongest possible adaptive capacity to environmental challenges. The benefit to immunity as shown here confirms the adaptive value of allelic diversity key to Darwin's theory. The absence of a link between diseases and MAs of slow evolutionary rate confirms the proven virtues of the neutral framework as applied to neutral sequences yet to reach optimum diversity. However, the existing framework fails to recognize another related intuition key to the MGD hypothesis that excess entropy/noises would be deleterious even though limited amount could be neutral or beneficial.

Most mutations are random and hence disorderly in origin (regardless whether the mutations may benefit a specific gene function), which can only be expected to be disruptive to the internal integrity of a system of great order. Any positively selected beneficial mutation therefore must be also at some point under negative selection because as a disorderly error it would contribute to the accumulation of entropy. This intuition is strongly supported by the data here. The existing framework also considers any observed genetic variations to be nowhere near an optimum level if granting any such level at all. It fails to see that if genetic variation in the form of limited amount of MAs is adaptive as in immunity, it would be *quickly* positively selected to reach a level where it cannot go beyond due to negative selection.

The effects of excess MAC in humans, both yes (such as T2D) and no (blood pressure), are remarkably similar to those in model organisms, suggesting that the link between excess MAC and diseases in humans is causal since it can be independently replicated in model organisms. This is all the more remarkable as the MAs were identified differently and independently in humans and model organisms, indicating again that MA amounts matter more than the specific function of individual MA. Also reassuring is that many MAC-linked traits were independently observed in different panels of RILs. Taken together, these independent and mutually supportive results confirm the power and validity of the experimental approaches employed here.

The results here suggest insights into the pathogenesis of certain diseases. It is well established that accumulation of somatic mutations causes cancer, although most such mutations are assumed to be neutral or “passengers” rather than “drivers” (Vogelstein, et al., 2013; Wood, et al., 2007). The results here that individuals with more germline SNPs or MAC have higher lung cancer incidences suggests an oncogenic role for most presumed neutral mutations and may explain the well-known variation in humans in lung cancer risk when exposed to the same pathogenic factors such as cigarette smoking. This makes good sense since cells with more random variations or SNPs should have more entropy, which would make growth control less precise and stable. Alcohol addiction in humans is associated with lower initial sensitivity to alcohols/drugs and strong alcohol and sweet preference and consumption (Crabbe, 2002). Strains with high MAC showed these phenotypes and may thus serve as good models of human alcoholism, although more future studies are needed. High MAC in mice were linked with increase in resistin and insulin level and decrease in IL-17 level. Such alterations have been implicated in mouse and human obesity and T2D and may mediate in part the pathogenic pathways in T2D associated with MAs (Steppan, et al., 2001; Zuniga, et al., 2010). If these biochemical changes could be confirmed by future studies, mice with high MAC may serve as useful models of obesity and T2D. Also, T2D patients are known to have higher risk of PD (Hu, et al., 2007), which is associated with loss of dopamine secreting neurons in the substantia nigra. Greater than half of PD patients exhibit abnormal glucose tolerance or diabetes (Boyd, et al., 1971; Lipman, et al., 1974). The finding of MAC association with both glucose and dopamine level, if confirmed by future studies, may help understand the common genetic mechanisms for T2D and PD.

MAC may cause disease susceptibility by making inherent traits less stably maintained or inherited during cell division (Zhu, et al., 2013). But the mechanism of action of MAC in complex diseases/traits may be hard to delineate precisely and usefully, since the defining characteristic of complexity may be the breakdown of causality. As simply put by Goldenfeld and Woese, “complex systems are ones for which observed effects do not have uniquely definable causes, due to the huge nature of the phase space and the multiplicity of paths.” (Goldenfeld and Woese, 2011). Thus, holistic system or architectural plan approaches may be more productive in studying MAC action than old fashioned ways of focusing on individual building blocks or genes.

The number of MAC-linked traits found in this analysis of ~4700 traits in genetic reference populations may be only a conservative minimum since most of these traits were only assayed in few strains, which cannot meet sampling size requirements for significant results. Also, if the SNPs used for calculating MAC were not truly a random sampling of the genome, they could miss certain traits. In the absence of complete SNP genotyping data representing all haplotypes, it may be premature to identify trait-specific set of MAs since such set is inherently incomplete. But even an incomplete set may still prove useful for certain practical purposes such as trait/disease risk predictions. Future studies using the methods here or improved versions of them may uncover more MAC-associated traits and identify trait/disease specific set of MAs useful for diagnosis.

**Materials and methods:**

**MAC calculation and statistical methods:**

Disease SNP GWAS datasets were downloaded from European Genome-phenome Archive (EGA) and database of Genotypes and Phenotypes (dbGaP). Duplicated individual datasets were excluded from the analysis. For certain large datasets such as PD and major depression, the data were randomly split into two cohorts to mimic two independent studies. SNP data for the genetic reference populations were obtained from the literature and public databases. All analyses were done with autosomal SNPs. Phenotype and gene expression data were from the literature, GeneNetwork, and Gene Expression Omnibus.

The MAF of each SNP in a RIL panel or a control cohort was calculated by PLINK and SNP Tools for Microsoft Excel (Chen, et al., 2009; Purcell, et al., 2007). The number of strains in each RIL panel is given in Supp. Table S1. From such MAF data, we obtained the MA set, which excluded non-informative SNPs with MAF = 0 in both cases and controls or in a RIL panel and with MAF = 0.5 in controls or a RIL panel. The MA set was equivalent to an imagined individual who is homozygous for all the MAs of informative SNPs analyzed.

Three SNP parameters related to nucleotide diversity were scored using a novel software Nucleotide Diversity 1 (ND1) developed for this study, including distance to the MA set, Het number, and pairwise genetic distance (PGD). Every non-repetitive pair within a population was scored to produce the average PGD. The ND1 software measures genetic distance between two individuals by the number of mismatched SNPs. For homozygous (Hom) vs Hom mismatch, a difference of 1 was scored. For Hom vs Het, a difference of 0.5 was scored. For Het vs Het, a difference of 0.5 was scored since half of such cases are expected to be A/B vs A/B with a difference of 0 whereas the other half are expected to be A/B vs B/A with a difference of 1. We assumed that on a genome wide scale, the number of A/B vs A/B match due to IBD (identical by descent) is similar to the number of A/B vs B/A mismatch. We verified this approach by comparing the PGD in X chromosome for CEU females vs CEU males using HapMap SNP data and found them to be similar as expected. In contrast, a software based on IBS (identical by status) such as PEAS that scores Het vs Het as 0 showed the males to have much greater PGD in X than females (Xu, et al., 2010). For the missing genotypes N/N, N/N vs Hom was scored as 0 and N/N vs Het as 0.5.

From the distance to the MA set which was the number of mismatched SNPs between an individual and the MA set, we obtained the amount of MAs each individual carries by subtracting the distance from the total number of SNPs scanned. The MAC of a strain was calculated by dividing the number of MAs carried by the strain by the number of total SNPs scanned. The effect of non-informative N/N genotypes has been taken into account or corrected for calculating the MAC, as wells as for calculating the distance to the MA set and the number of Het. For the distance to the MA set, the formula for N/N correction is: Corrected distance Y = Pre-correction distance + # N/N x (Y/# total SNPs).

The correlation between genotype and phenotype was analyzed by linear and multivariate regression analysis using GraphPad Prism 5 and InStat3 and SAM. For multivariate regression analysis of the 3664 traits in BXD panel, most traits were unsuitable for analysis because of missing data. After removing these, there were 21 traits left, from which 13 were filtered out because of non-independent nature based on multivariate analysis. The remaining 9 traits were then analyzed by multivariate regression using InStat3. Other statistical methods used include Student's t test, two tailed, chi-square test, two tailed, linear and multivariate regression, and Pearson/Spearman correlation analysis.

**Identification of trait specific set of MAs**

To confirm among the MA-linked traits that correlated traits share more MAs than non-correlated traits, we examined four traits shown in Table S2, blood ethanol concentration (BEC, trait 3), its strongly correlated trait resistin level (trait 11), and its two non-correlated traits pain response (trait 4) and open field rearing behavior (trait 5), in order to identify a BEC specific set of SNPs or MAs. From 51469 SNPs originally used for calculating MAC for the BXD panel, we identified 34709 SNPs using 61 strains with BEC data, with each SNP having more MAs in the bottom one third of the panel with the lowest BEC relative to the top one third with the highest

BEC. We did this by converting each genotype of a SNP in a strain into a score of 0 for hom MA, 0.5 for het, and 1 for hom major allele. These MAs were then ranked by *p* value from t test on their different enrichment in the bottom one third versus the top one third of the panel, and divided into several groups, including groups with *p* <0.05, 0.05-0.11, 0.11-0.2, <0.2, 0.2-0.5, <0.5, 0.5-0.75, <0.75, 0.75-1, and 0-1. We used each group of SNPs or MAs to calculate the MAC of each strain, which were then used to correlate with the 4 traits for a panel of 24 strains with each having data for all 4 traits. The group of 30336 SNPs with P < 0.75 was found to have the best BEC specificity.

**Identification of non-synonymous SNPs located in the slowest evolving genes**

To obtain non-synonymous SNPs located in the slowest evolving genes, we collected the whole genome protein data of *Homo sapiens* (version 36.3) and *Macaca mulatta* (version 1) from the NCBI ftp site (ftp://ftp.ncbi.nih.gov/genomes/) and then compared the human protein to the monkey protein using local BLASTP program at a cut-off of 1E-10. We only retained one human protein with multiple isoforms and chose the monkey protein with the most significant E-value as the orthologous counterpart of each human protein. The aligned proteins were ranked by percentage identities. Proteins that show the highest identity between human and monkey were considered the slowest evolving (including 83 genes > 200 amino acid in length with 100% identity and 74 genes > 1105 amino acid in length with 99% identity between monkey and human), which additionally showed fewer number of coincident or overlapped substitutions than fast evolving genes among different lineages. We downloaded the HapMap coding SNPs data from the dbSNP database (Build 130; ftp://ftp.ncbi.nih.gov/snp/), including the ancestral allele information, the gene locations of all SNPs and the allele frequencies of SNPs in four populations CEU, CHB, JPT and YRI. We thus obtained a list of 414 non-synonymous SNPs located in the slowest evolving proteins.

**Animal experiments:**

**Ethics Statement:** The study performed animal experiments and the animals' care was in accordance with institutional guidelines. The Institutional Animal Care and Use Committee of the Central South University has approved this study.

For brood size measurement, all lines were synchronized by transferring five adult nematodes to fresh dishes and allowing them to lay eggs for 3-4 h, after which the nematodes were removed. Twenty L4 individuals from each line were picked into 20 dishes and were allowed to lay eggs each day into a new dish for a total 8 days or until no more eggs were laid. The eggs in each dish were allowed to develop for 2 days before being counted.

See Extended experimental procedures for experiments on immune responses and high fat diet-induced obesity in BXD mice.

**Acknowledgements:**

We thank E. Andersen, L. Kruglyak, A. Bendesky, P. McGrath, C. Bargmann, E. Chesler, M. Rockman, R. Brem, E. Smith, L. Xia, J. Li, B. Chen, L. Zhou, Y. Zheng, R. Mott, X. Zhou, and R. Williams for research materials or technical assistance, and A. Bendesky and C. Bargmann for critical reading of the manuscript. We thank the two reviewers for their constructive comments that helped to strengthen our article. Supported by the National Natural Science Foundation of China grant 81171880 and the National Basic Research Program of China grant 2011CB51001 (S.H.), and the GeNeSys Consortium (O.G. and E.M).

**Figure Legends:**

**Figure 1. Distribution profile of MAC of each strain in a panel of genetic reference population.** A. RIL strains from the Kruglyak laboratory separated by the *npr-1* F215V mutation into the HW and N2 types. B. RIL strains from the Kammenga laboratory. C. Budding yeast segregant panel. D. *D. melanogaster* inbred strain panel derived from randomly selected individuals in the wild. E. Mouse RIL panel from Collaborative Cross at 7$^{th}$ generation of inbreeding. F. Mouse BXD RIL panel.

**Figure 2. MAC on reproductive fitness in *C. elegans*.** A. Brood size in Kruglyak RIL strains with HW *npr-1* genotype. B. Brood size in Kruglyak RIL strains with N2 *npr-1* genotype.

**Figure 3. MAC on ethanol traits in BXD mice.** A. Maximum threshold to ethanol induced ataxia. B. Blood ethanol concentration in males.

**Figure 4. MAC on tumorigenesis in mouse RILs.** A. The number of lung tumors induced by urethane in *kras2* wild type AXBBXA strains. B. The number of lung tumors induced by urethane in *kras2* mutant AXBBXA strains. Also shown are average tumor values of top and bottom half in MAC and t test *p* values.

**Figure 5. MAC on obesity and PD related traits in rat RILs.** A. Glucose concentration in 10 week old BXHHXB male rats fed a diet with 60% fructose from 8 weeks to 10 weeks. B. Serum dopamine level in 6 week old male BXHHXB rats.

**Figure 6. Purifying selection of both high and low MAC.**

**Table Legends:**

**Table 1.** MAC on immune responses in mouse RIL panels.

**Table 2.** Distance to the MA set, heterozygosity, pairwise genetic distance (PGD), and correlation between MAC and case status. [a] Shaded numbers indicate greater MA content or nucleotide diversity. Numbers represent ratio of the distance to the MA set (or Het# or PGD) over the number of SNPs analyzed. [b] t-test, *p* value, two tailed. [c] ~0 means $p < 2.2E-16$.

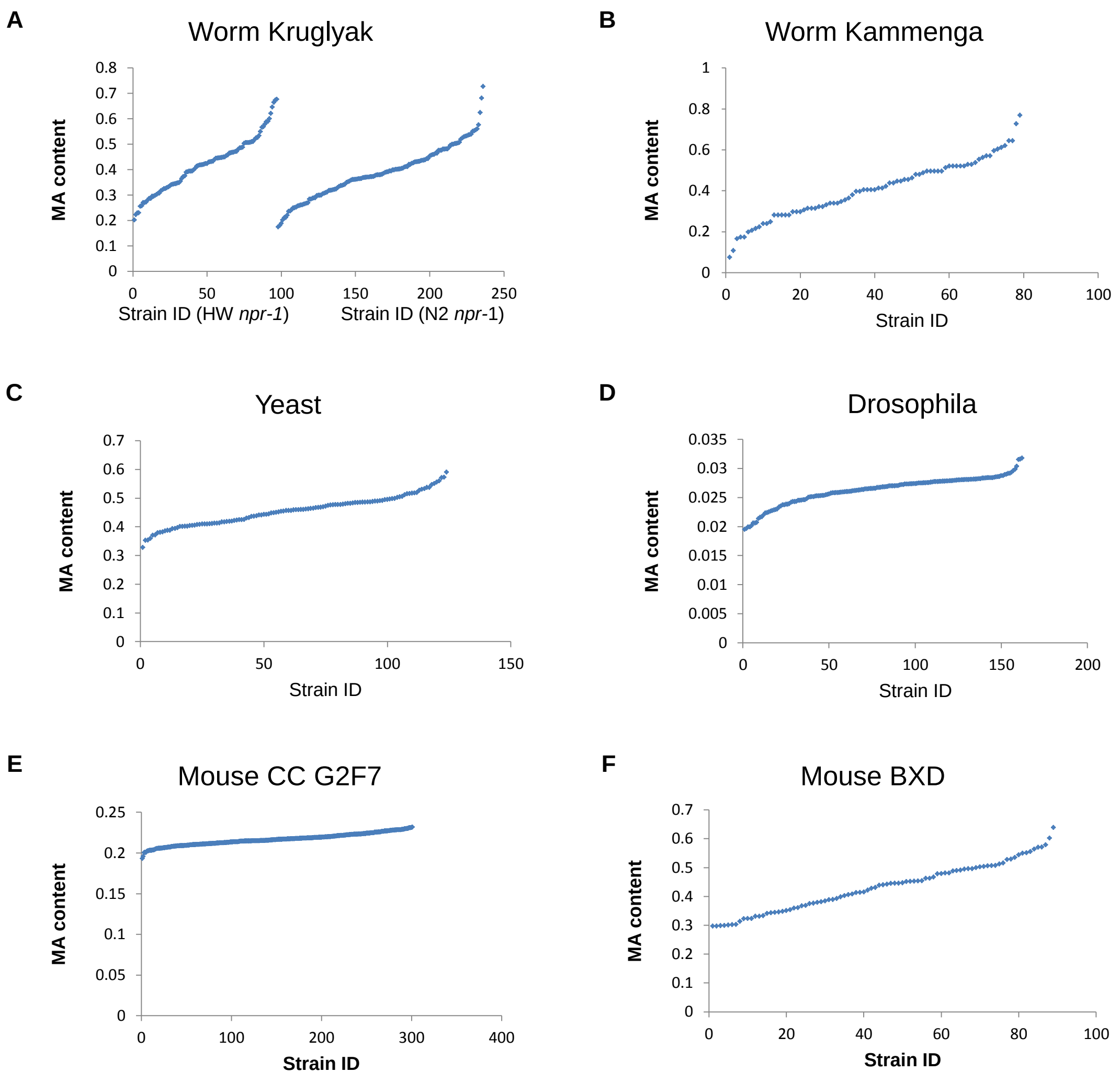

Fig 1

A

Spearman r = -0.44 $p$ = 0.004

Pearson r = -0.45 $p$ = 0.003

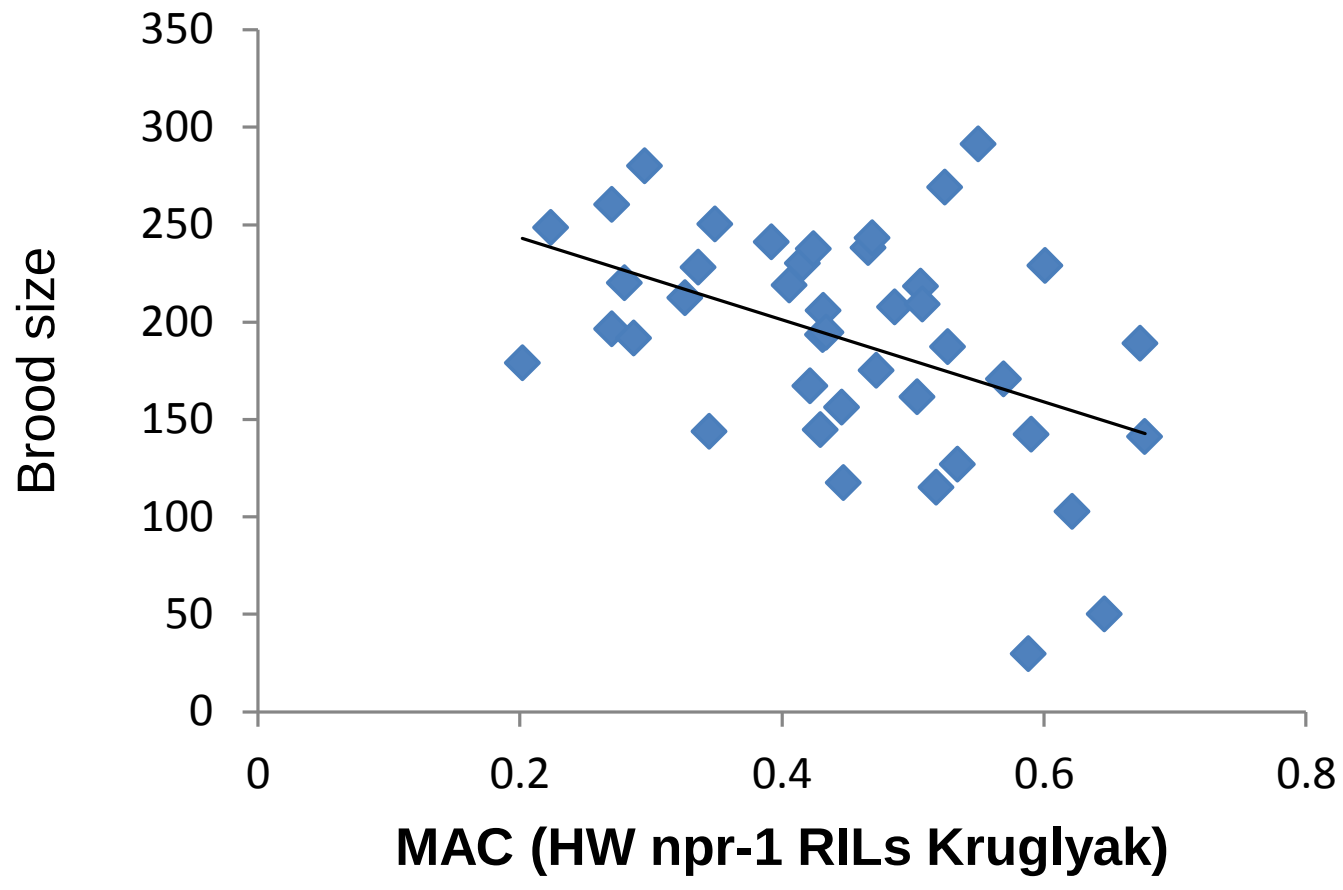

**B**

Spearman r = -0.30 $p$ = 0.02

Pearson r = -0.26 $p$ = 0.04

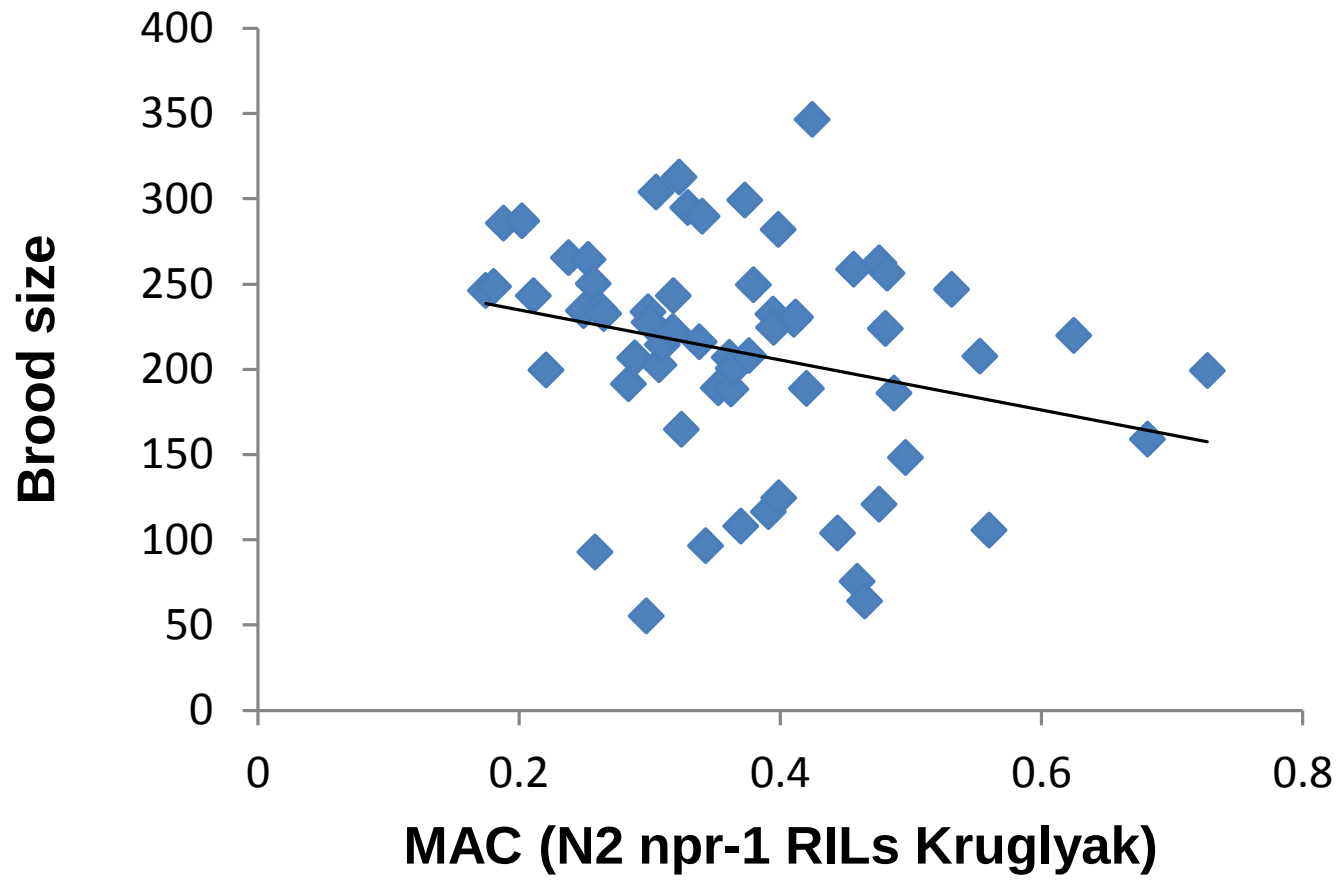

Fig 2

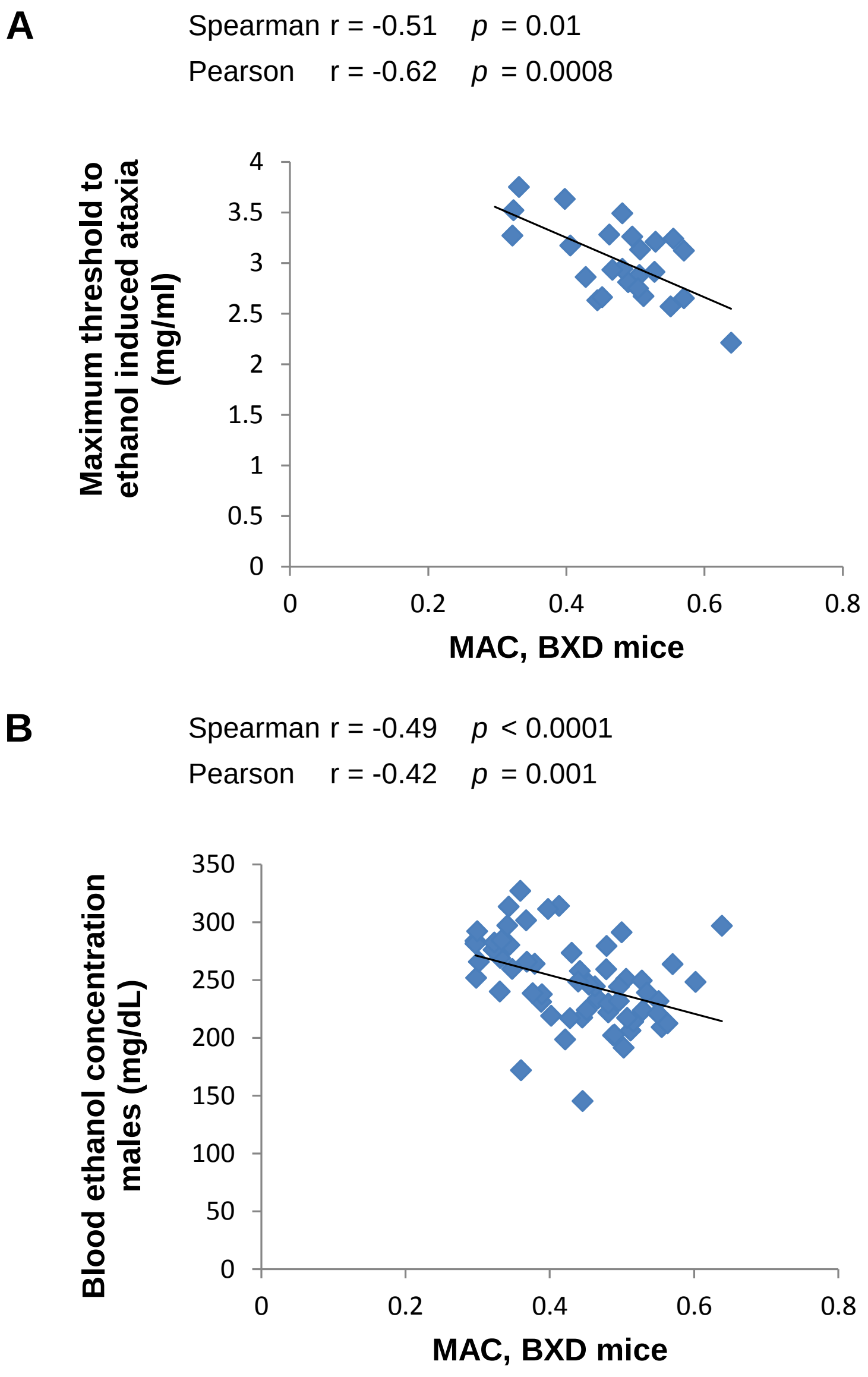

Fig. 3

A

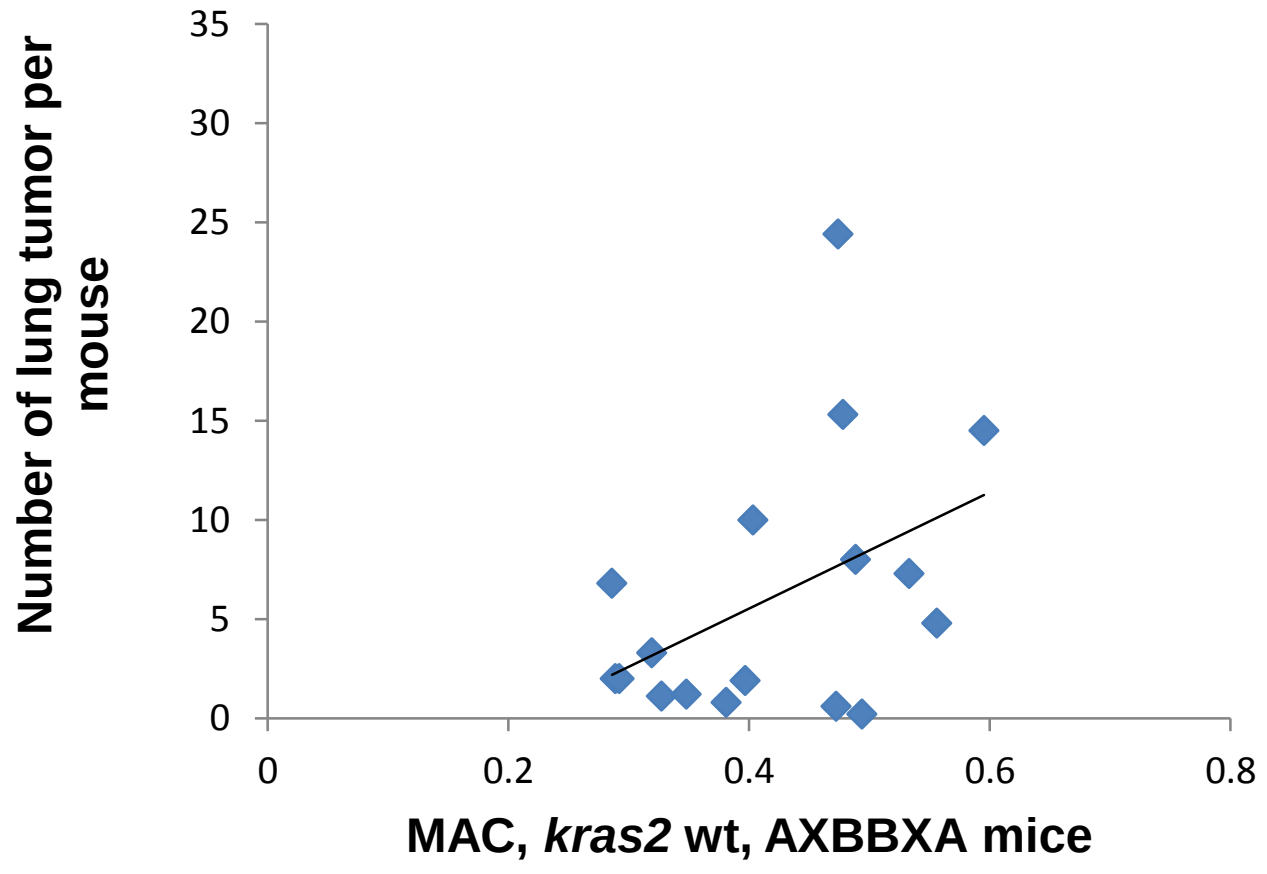

| Strains | Average |
|---|---|
| Low MA (8) | 2.4 ± 1.9 |
| High MA (9) | 9.4 ± 7.7 |
| *p* (t test) | 0.02 |

B

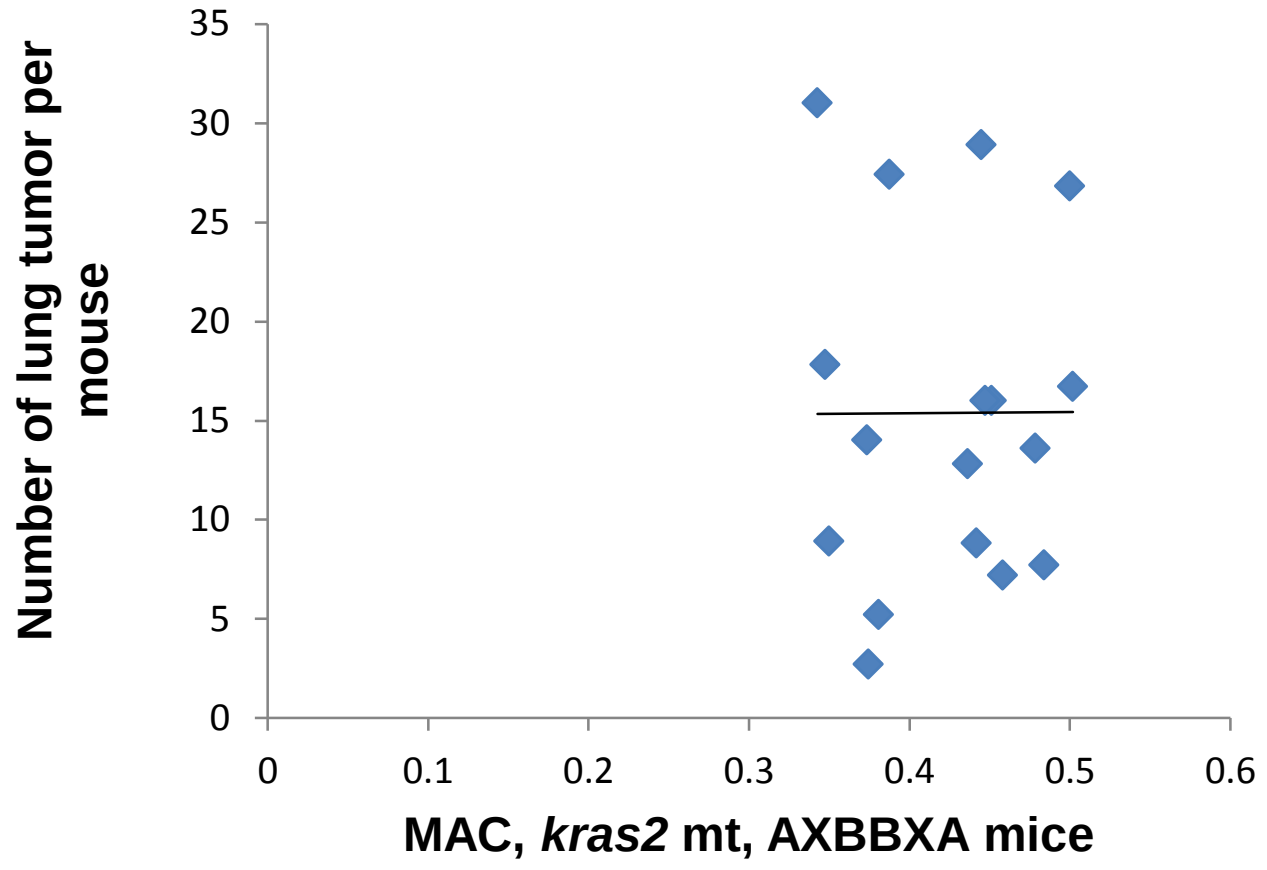

| Strains | Average |
|---|---|
| Low MA (8) | 15.0 ± 10.1 |
| High MA (9) | 15.7 ± 7.8 |
| *p* (t test) | ns |

Fig 4

**A**

Spearman r = 0.62 $p$ = 0.0009
Pearson r = 0.67 $p$ = 0.0002

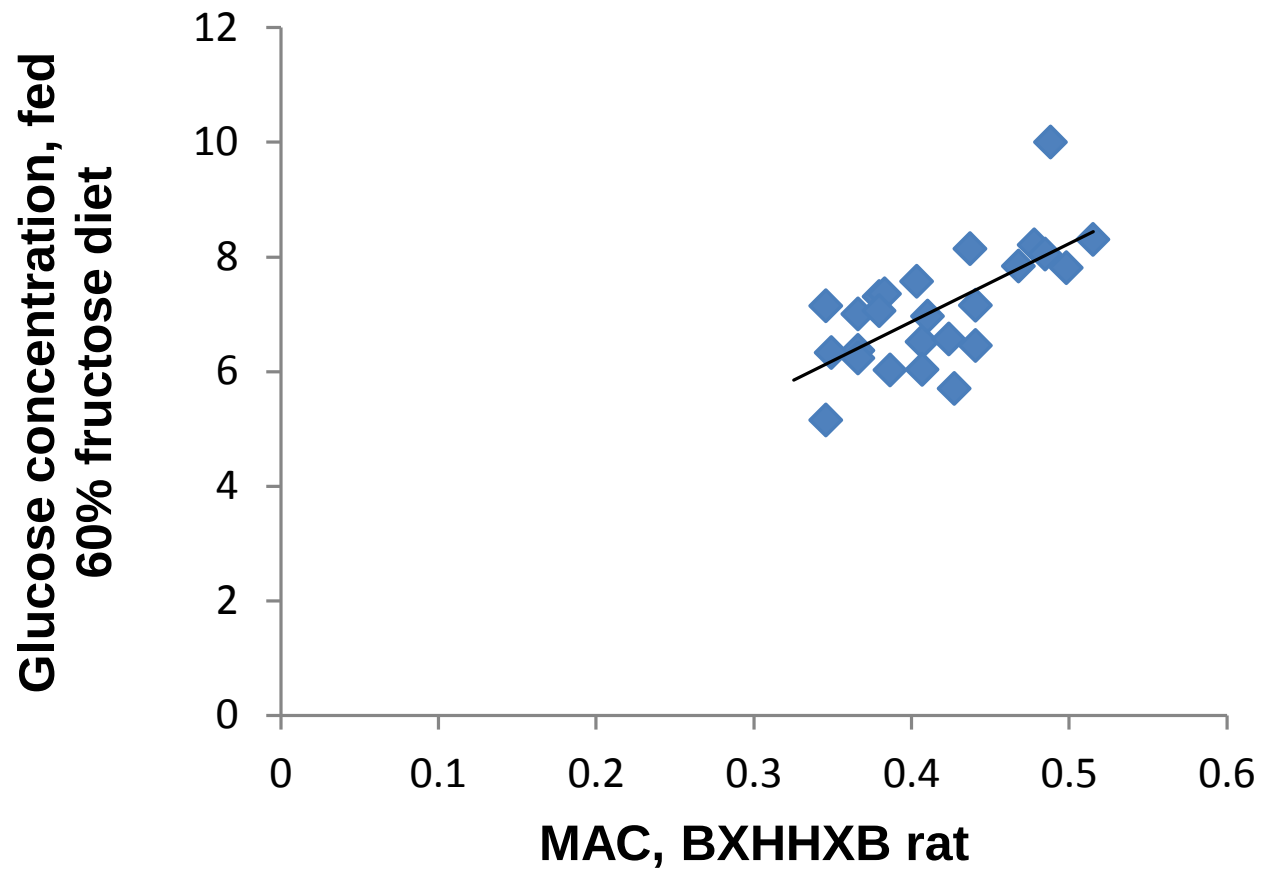

**B**

Spearman r = -0.45 $p$ = 0.02
Pearson r = -0.41 $p$ = 0.03

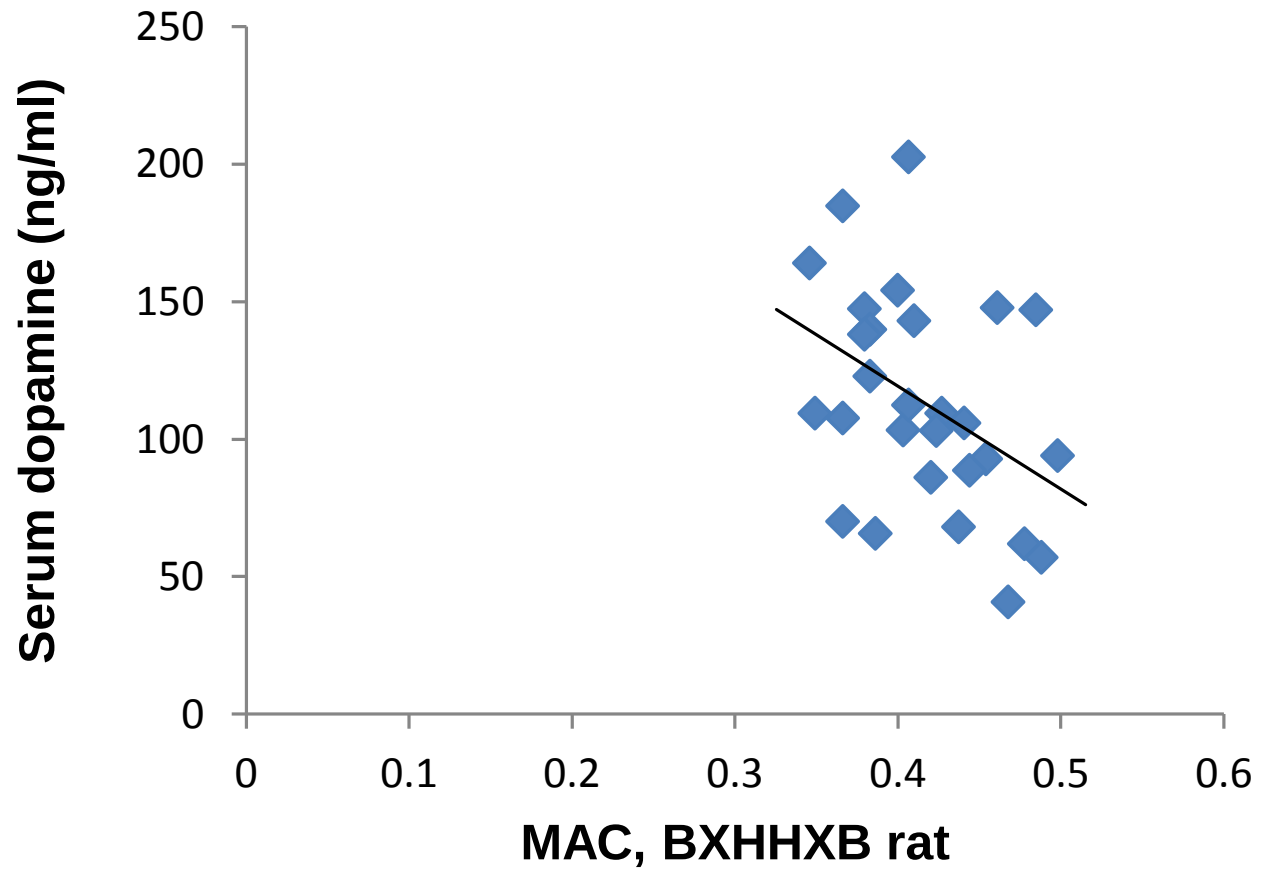

Fig. 5

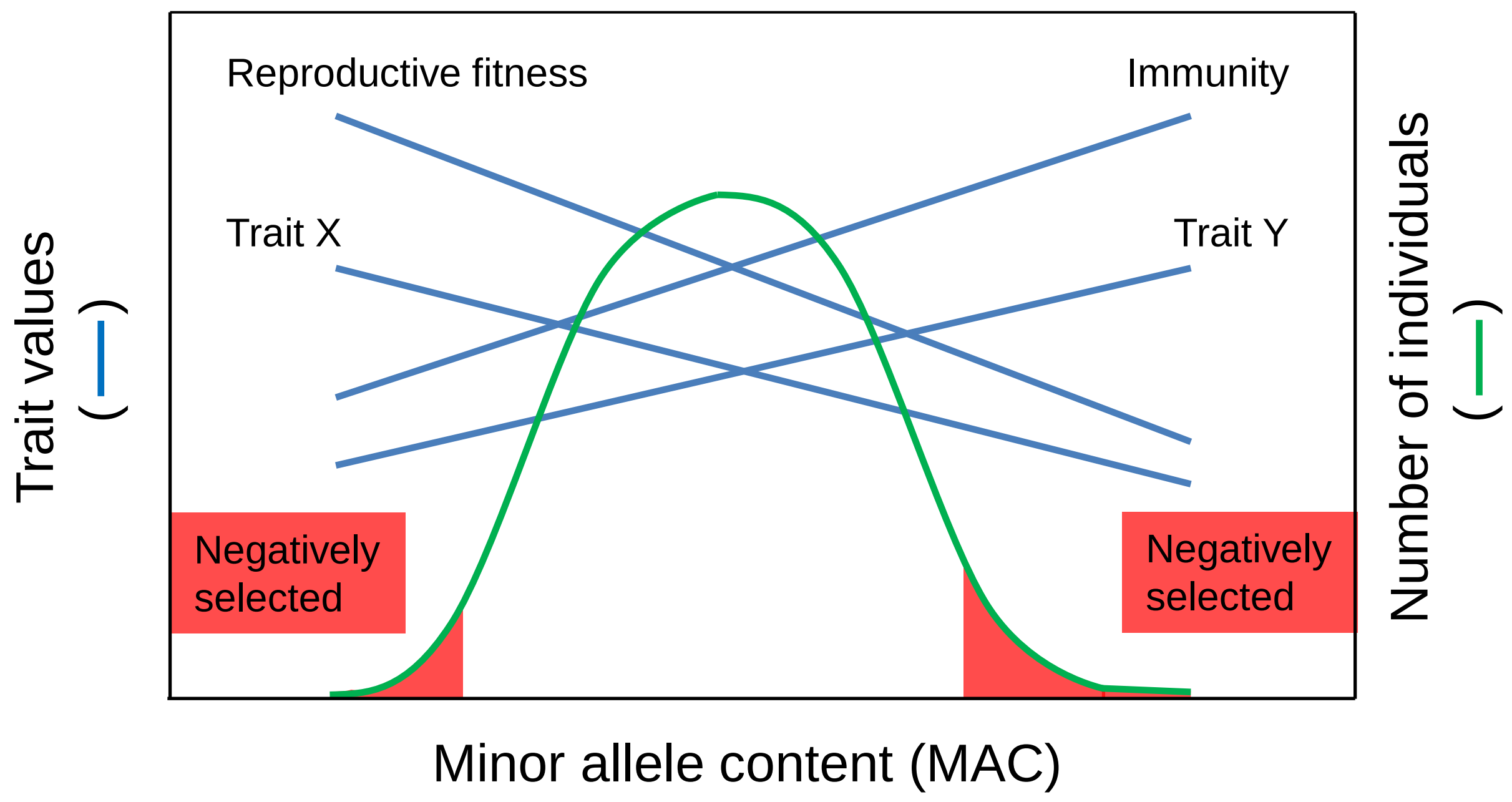

**Table 1. MAC on immune responses in mouse RIL panels**

| Mice | GN ID[a] | Sample # | Trait description | Pearson | | Spearman | |
|---|---|---|---|---|---|---|---|
| | | | | r | *p* | r | *p* |
| BXD | 13969 | 24 | CFU in liver 48h post i.v. *S. aureus* infection | -0.57 | 0.003 | -0.66 | 0.0005 |
| | 14313 | 6 | White blood cell count after *C. albicans IV* infection | 0.93 | 0.01 | 0.83 | 0.04 |
| | 10779 | 18 | IgG1 anti-cF.IX (coagul. Factor IX) after cF.IX injection | 0.6 | 0.01 | 0.48 | 0.04 |
| | 10695 | 21 | Pulmonary granulomatous inflammation by BCG | -0.47 | 0.03 | -0.39 | ns |
| | 12668 | 33 | Formation of dermal lesions, ECTV footpad | -0.36 | 0.04 | -0.32 | ns |
| | 10663 | 20 | Cytotoxicity in spleen T cells post AdLacZ i.v injection | -0.43 | ns | -0.55 | 0.01 |
| | 10806 | 25 | Mortality after i.p. *C. psittaci* infection | -0.39 | 0.05 | -0.43 | 0.03 |
| BXH | 10115 | 10 | Survival times of allograft | -0.68 | 0.02 | -0.6 | 0.04 |

[a]GeneNetwork identification number.

**Table 2. Distance to the MA set, heterozygosity, pairwise genetic distance (PGD), and correlation between MAC and case status.**

| | All types of SNPS | | | | | | | | | Non-syn SNPs | | |
|---|---|---|---|---|---|---|---|---|---|---|---|---|
| | | | | | | | Spearman | | | | | |
| Disease | Dist. MA | stdev | Het | stdev | PGD | stdev | r | *p* | SNP # | Dist. MA | SNP# | Sample # |
| WTCCC T2D | | | | | | | | | | | | |
| case | 0.7958[a] | 0.0083 | 0.2780 | 0.0095 | 0.2774 | 0.0089 | 0.09 | <0.0001 | 485959 | 0.8288 | 23 | 1999 |
| control | 0.7963 | 0.0022 | 0.2769 | 0.0035 | 0.2770 | 0.0024 | | | | 0.8263 | | 3004 |
| *p*[b] | 0.008 | | 2.0E-08 | | ~0[c] | | | | | ns | | |
| dbGAP T2D European American (EA) female, phs000091.v2.p1 | | | | | | | | | | | | |
| case | 0.7968 | 0.0041 | 0.2750 | 0.0031 | 0.2742 | 0.0025 | 0.05 | 0.002 | 830203 | 0.8815 | 49 | 1767 |
| control | 0.7976 | 0.0047 | 0.2751 | 0.0035 | 0.2743 | 0.0033 | | | | 0.8220 | | 1537 |
| *p* | 5.5E-08 | | ns | | ~0 | | | | | ~0 | | |
| dbGAP T2D EA male, phs000091.v2.p1 | | | | | | | | | | | | |
| case | 0.7926 | 0.0047 | 0.2804 | 0.0037 | 0.2805 | 0.0033 | ns | | 811523 | 0.8180 | 48 | 1115 |
| control | 0.7928 | 0.0048 | 0.2803 | 0.0037 | 0.2804 | 0.0033 | | | | 0.8154 | | 1277 |
| *p* | ns | | ns | | ~0 | | | | | 0.04 | | |
| dbGAP PD cohort1 EA, phs000196.v2.p1 | | | | | | | | | | | | |
| case | 0.7855 | 0.0024 | 0.2899 | 0.0028 | 0.2903 | 0.0025 | 0.14 | <0.0001 | 863755 | 0.8587 | 168 | 1000 |
| control | 0.7859 | 0.0026 | 0.2898 | 0.0024 | 0.2905 | 0.0081 | | | | 0.8576 | | 999 |
| *p* | 0.0006 | | ns | | ~0 | | | | | ns | | |
| dbGAP PD cohort2 EA, phs000196.v2.p1 | | | | | | | | | | | | |
| case | 0.7850 | 0.0017 | 0.29063 | 0.0029 | 0.2910 | 0.0018 | ns | | 861405 | 0.8594 | 169 | 1011 |
| control | 0.7854 | 0.0016 | 0.29064 | 0.0034 | 0.2908 | 0.0016 | | | | 0.8596 | | 993 |
| *p* | 6.1E-05 | | ns | | ~0 | | | | | ns | | |
| dbGAP PD H550,EA, phs000089.v3.p2 | | | | | | | | | | | | |
| case | 0.7604 | 0.0044 | 0.3175 | 0.0065 | 0.3236 | 0.0032 | 0.21 | <0.0001 | 544580 | 0.7754 | 84 | 672 |
| control | 0.7611 | 0.0041 | 0.3194 | 0.0052 | 0.3233 | 0.0024 | | | | 0.7746 | | 527 |
| *p* | 2.3E-05 | | 7.2E-09 | | ~0 | | | | | ns | | |
| dbGAP Major depression cohort 1, phs000020.v2.p1 | | | | | | | | | | | | |
| case | 0.7468 | 0.0086 | 0.3420 | 0.0054 | 0.3419 | 0.0038 | 0.13 | <0.0001 | 450394 | 0.8336 | 109 | 1000 |
| control | 0.7476 | 0.0059 | 0.3418 | 0.0037 | 0.3413 | 0.0023 | | | | 0.8354 | | 1000 |
| *p* | 0.005 | | ns | | ~0 | | | | | ns | | |
| dbGAP Major depression cohort 2, phs000020.v2.p1 | | | | | | | | | | | | |
| case | 0.7470 | 0.0129 | 0.3419 | 0.0075 | 0.3414 | 0.0047 | 0.15 | <0.0001 | 450338 | 0.8317 | 109 | 813 |
| control | 0.7476 | 0.0064 | 0.3416 | 0.0041 | 0.3413 | 0.0028 | | | | 0.8350 | | 911 |
| *p* | 0.006 | | ns | | ~0 | | | | | 0.006 | | |
| WTCCC Bipolar disorder | | | | | | | | | | | | |
| case | 0.7946 | 0.0049 | 0.2794 | 0.0063 | 0.2791 | 0.0052 | 0.08 | <0.0001 | 482969 | 0.8409 | 25 | 1998 |
| control | 0.7950 | 0.0022 | 0.2786 | 0.0035 | 0.2787 | 0.0024 | | | | 0.8404 | | 3004 |
| *p* | 8.7E-05 | | 7.3E-08 | | ~0 | | | | | ns | | |
| dbGAP Bipolar disorder EA, phs000017 | | | | | | | | | | | | |
| case | 0.7956 | 0.0087 | 0.2783 | 0.0052 | 0.2740 | 0.0028 | 0.14 | <0.0001 | 845359 | 0.8644 | 47 | 1102 |
| control | 0.7961 | 0.0073 | 0.2773 | 0.0043 | 0.2741 | 0.0023 | | | | 0.8635 | | 1081 |
| *p* | 5.0E-10 | | ns | | ~0 | | | | | ns | | |

**Table 2. Distance to the MA set, heterozygosity, pairwise genetic distance (PGD), and correlation between MAC and case status.(Continued)**

| | | | | | | | | | | | | |
|---|---|---|---|---|---|---|---|---|---|---|---|---|
| dbGAP GAIN Schizophrenia EA, phs000021.v2.p1 | | | | | | | | | | | | |
| case | 0.7988 | 0.0071 | 0.2732 | 0.0045 | 0.2705 | 0.0023 | ns | | 858419 | 0.8254 | 50 | 1214 |
| control | 0.7992 | 0.0071 | 0.2730 | 0.0040 | 0.2693 | 0.0125 | | | | 0.8239 | | 1442 |
| *p* | ns | | 0.01 | | ~0 | | | | | ns | | |
| dbGAP nonGAIN Schizophrenia EA, phs000167.v1.p1 | | | | | | | | | | | | |
| case | 0.7987 | 0.0074 | 0.2734 | 0.0059 | 0.2707 | 0.0038 | 0.07 | 0.007 | 861559 | 0.8235 | 52 | 1034 |
| control | 0.7990 | 0.0077 | 0.2732 | 0.0047 | 0.2704 | 0.0028 | | | | 0.8276 | | 952 |
| *p* | ns | | ns | | ~0 | | | | | ns | | |
| dbGAP Lung cancer H550/Ilmn, phs000336.v1.p1, | | | | | | | | | | | | |
| case | 0.7602 | 0.0019 | 0.32375 | 0.0023 | 0.3240 | 0.0017 | 0.14 | <0.0001 | 542243 | 0.7767 | 84 | 776 |
| control | 0.7605 | 0.0017 | 0.32376 | 0.0028 | 0.3242 | 0.0015 | | | | 0.7747 | | 845 |
| *p* | 0.001 | | ns | | ~0 | | | | | ns | | |
| dbGAP Lung cancer H610/Ilmn, phs000336.v1.p1 | | | | | | | | | | | | |
| case | 0.7888 | 0.0331 | 0.2817 | 0.0493 | 0.3187 | 0.0024 | 0.12 | <0.0001 | 571575 | 0.7888 | 90 | 3099 |
| control | 0.7910 | 0.0330 | 0.2780 | 0.0507 | 0.3181 | 0.0026 | | | | 0.7910 | | 1925 |
| | 3.7E-15 | | 4.1E-08 | | ~0 | | | | | 8.0E-06 | | |
| dbGAP Breast cancer, phs000147.v1.p1 | | | | | | | | | | | | |
| case | 0.7601 | 0.0046 | 0.32379 | 0.0032 | 0.3226 | 0.0020 | 0.044 | <0.05 | 531623 | 0.7748 | 83 | 1069 |
| controal | 0.7604 | 0.0042 | 0.32381 | 0.0027 | 0.3226 | 0.0018 | | | | 0.7771 | | 1083 |
| *p* | ns | | ns | | ns | | | | | ns | | |
| dbGAP Alcohol addiction, phs000092.v1.p1, EA cohort | | | | | | | | | | | | |
| case | 0.7859 | 0.0040 | 0.2888 | 0.0039 | 0.2889 | 0.0035 | 0.042 | <0.05 | 959085 | 0.8708 | 189 | 1167 |
| control | 0.7862 | 0.0037 | 0.2888 | 0.0037 | 0.2887 | 0.0031 | | | | 0.8710 | | 1366 |
| *p* | ns | | ns | | ns | | | | | ns | | |
| WTCCC T1D | | | | | | | | | | | | |
| case | 0.7947 | 0.0025 | 0.2669 | 0.0034 | 0.2787 | 0.0028 | 0.06 | <0.0001 | 478370 | 0.8416 | 25 | 2000 |
| control | 0.7949 | 0.0022 | 0.2789 | 0.0034 | 0.2788 | 0.0024 | | | | 0.8402 | | 3004 |
| *p* | 0.022 | | ~0 | | ~0 | | | | | ns | | |
| WTCCC Rheumatoid arthritis | | | | | | | | | | | | |
| case | 0.7958 | 0.0024 | 0.2776 | 0.0089 | 0.2773 | 0.0105 | 0.036 | 0.01 | 481944 | 0.8516 | 27 | 1999 |
| control | 0.7964 | 0.0022 | 0.2766 | 0.0035 | 0.2767 | 0.0024 | | | | 0.8521 | | 3004 |
| *p* | 0.009 | | 4.1E-07 | | ~0 | | | | | ns | | |
| WTCCC Inflammatory bowel disease or Crohn's disease | | | | | | | | | | | | |
| case | 0.7954 | 0.0135 | 0.2773 | 0.0118 | 0.2781 | 0.0144 | 0.085 | <0.0001 | 483234 | 0.8463 | 26 | 2005 |
| control | 0.7970 | 0.0022 | 0.2758 | 0.0035 | 0.2759 | 0.0024 | | | | 0.8464 | | 3004 |
| *p* | 4.7E-07 | | 7.4E-08 | | ~0 | | | | | ns | | |
| WTCCC Coronary artery disease | | | | | | | | | | | | |
| case | 0.79448 | 0.0024 | 0.2795 | 0.0052 | 0.2791 | 0.0026 | ns | | 477416 | 0.8015 | 25 | 1988 |
| control | 0.79447 | 0.0022 | 0.2792 | 0.0035 | 0.2793 | 0.0024 | | | | 0.8002 | | 3004 |
| *p* | ns | | 0.007 | | ~0 | | | | | ns | | |
| WTCCC Hypertension | | | | | | | | | | | | |
| case | 0.7936 | 0.0014 | 0.2809 | 0.0049 | 0.2804 | 0.0015 | ns | | 479524 | 0.8342 | 24 | 2001 |
| control | 0.7935 | 0.0022 | 0.2806 | 0.0035 | 0.2807 | 0.0024 | | | | 0.8336 | | 3004 |
| *p* | ns | | 0.03 | | ~0 | | | | | ns | | |

[a] Shaded numbers indicate greater MA content or nucleotide diversity. Values represent ratio of the distance to the MA set (or het# or PGD) over the number of SNPs analyzed.
[b] t test *p* value.
[c] ~0 means $p < 2.2E-16$.

# Methods for scoring the collective effect of SNPs: Minor alleles of common SNPs quantitatively affect traits/diseases and are under both positive and negative selection

**Supplementary Materials:**

**Supplementary Figures:**

**Supp. Figure S1. Population distribution bell curve of MAC in various RIL panels.**

**Supp. Figure S2. MAC value is independent of SNP numbers or random selection of SNPs.** A. The correlation curve between MAC of each BXD strain calculated from ~51K SNP and that from 1K SNP set 1 randomly selected from the ~51K SNP. B. The correlation curve between MAC of each BXD strain calculated from ~51K SNP and that from 1K SNP set 2 randomly selected from the ~51K SNP with no overlap with set 1.

**Supplementary Tables:**

**Supp. Table S1. MAC values of RIL strains or segregants**

**Supp. Table S2. Traits repeatedly linked (or not) with higher MAC.** [a]Reproductive fitness includes uterus horn length (BXD), brood or litter size (CC mice and worm), and fetal weights in left horn of uterus (BXHHXB). [b]Alcohol sensitivity was assayed by distance traveled after ethanol

injection. [c]Cancers include DEN induced liver tumors (BXD), urethane induced lung tumors with wild type *kras2* (AXBBXA), and virus induced lymphomas (CXB).

**Supp. Table S3. Correlation between traits and MAC values in BXD mice.** Significant associations are indicated by yellow color in column CW and CY.

**Supp. Table S4. Significant correlations among selected traits linked with higher MAC in BXD mice.** Significance scores from Pearson correlation analysis with $p$ <0.05, 0.01, and 0.001 represented by *, **, and ***, respectively.

**Supp. Table S5. Multivariate regression analysis.**

**Supp. Table S6. MA correlation with yeast growth in the presence of a compound.** [a]FDA-approved drugs in italics.

**Supp. Table S7. Correlations between higher MAC and clinical traits.**

**Extended experimental procedures:**

Naive inbred, specific pathogen-free (SPF status), 8- to 12-week-old BXD RIL animals as well as the parental C57BL/6J and DBA/2J were used for the study of immune response to *S. aureus.* Animals were housed at Harlan (the Netherlands). Mice were maintained under standard conditions and according to institutional guidelines. All experiments were approved by the appropriate ethical board (Niedersächsisches Landesamt für Verbraucherschutz und Lebensmittelsicherheit, Oldenburg, Germany). Mice were inoculated with 4 x $10^7$ cfu of *S. aureus* in 0.2 ml of PBS via a lateral tail vein. For determination of bacterial loads (cfu), infected mice were sacrificed by $CO_2$ asphyxiation at 48h post infection and the amount of bacteria determined by preparing liver homogenates in 5 ml of PBS and plating 10-fold serial dilutions on blood agar. Bacteria colonies were counted after incubation for 24 hours at 37°C.

BXD animals were used for ectromelia virus infection. 90 pfu of ectromelia virus was injected into one rear footpad of mice while under anesthesia. Animals were monitored daily for the formation of secondary dermal lesions over a two week period post infection. The presence of secondary lesions at any point over the 2 week period was scored as positive for the trait, the absence scored as negative for the trait.

For high fat diet experiment in mice, a total of 152 males of the parental strains C57BL/6J (B6) and DBA/2J (D2), $F_1$ offspring of the initial cross between B6 and D2 (B6D2F1), and 29 BXD RI strains (BXD 1, 2, 6, 8, 11, 14,15,16, 24a, 27, 31, 32, 33, 38, 39, 40, 42, 43, 51, 61, 62, 68, 69, 73, 75, 86, 87, 90, 96) were used in this study (Four to five males of each of the parental, $F_1$ and BXD strains were used). Mice were purchased from The Jackson Laboratory and were bred in the facility of the Neuro-Bsik consortium from the VU University Amsterdam, Netherlands. At the age of four weeks, mice were shipped to the mouse facility of the Department for Crop and Animal Sciences, Faculty of Agriculture and Horticulture at Humboldt-Universitat zu Berlin, Germany. Mice were maintained under conventional conditions and controlled lighting with a 12:12 hours light:dark cycle at a temperature of 22 ± 2 °C and a relative humidity of 65%. They were reared in groups of three to four individuals of the same sex in macrolon cages with a 350 $cm^2$ floor space (E. Becker & Co (Ebeco) GmbH, Castrop-Rauxel, Germany) and with bedding type S 80/150, dust-free (Rettenmeier Holding AG, Wilburgstetten, Germany). All individuals had *ad libitum* access to food and water.

Beginning at the age of 4 weeks, mice were fed a high-fat diet (HFD) (Ssniff® diet S8074-E010, Germany) until 20 weeks. The diet contained 20.7% crude protein, 25.1% crude fat, 5.0% crude fiber, 5.9% crude ash, 39.7% N-free extract, 20.0% starch, 17.5% sugar, vitamins, trace elements, amino acids, and minerals (19.1 MJ/kg metabolizable energy; thereof 45% energy from fat, 31% from carbohydrates, and 24% from proteins). The fat in the diet derived from coconut oil and suet.

Four to five males of each of the parental, $F_1$ and BXD strains were used for Attenuated Total Reflectance Fourier Transform Infrared Spectroscopy (ATR-FTIR) measurements. At 20 weeks, mice were fasted for two hours, anesthetized under isofluorane and decapitated using surgical scissors. After exsanguinations, reproductive fat pads (which was the epididymal adipose tissue), liver, and quadriceps muscle (comprised of *Musculus rectus femoris*, *Musculus vastus intermedius*, *Musculus vastus lateralis*, and *Musculus vastus medialis*) were dissected and weighed. All tissues were immediately frozen in liquid nitrogen and stored at -80 $^{o}$C until ATR-FTIR studies.

Phenotypes recorded:

Total fat weight at 42 days on high fat diet (g, 45% energy from fat) measure by Magnetic resonance interference (MRI)

Fat gain between 6 and 8 weeks on high fat diet (g, 45% energy from fat)

Total lean weight at 42 days on high fat diet (g, 45% energy from fat) measure by Magnetic resonance interference (MRI)

Lean gain between 6 and 8 weeks on high fat diet (g, 45% energy from fat)

Body weight at 28 days on high fat diet (g, 45% energy from fat)

Body weight gain between 5 and 6 weeks on high fat diet (g, 45% energy from fat)

Body length from the nose until the tail at 42 days

Serum glucose level at 70 days on high fat diet (g, 45% energy from fat)

Food Intake at 7 weeks

Food Intake between 7 and 10 weeks

Brain weight at 140 days on high fat diet (g, 45% energy from fat)

Weight of subcutaneous adipose tissue (upon the gluteal Mucsculus maximus between the legs, left and right of the tail) at 140 days on high fat diet (g, 45% energy from fat)

Weight of brown adipose tissue (located on the back, along the upper half of the spine and toward the shoulders) at 140 days on high fat diet (g, 45% energy from fat)

Serum level of leptin in males at 140 days on high fat diet (g, 45% energy from fat)

Unsaturated fat content in reproductive adipose tissue in males at 140 days on high fat diet measure by Fourier Transform Infrared Spectroscopy (FTIR)

**Supplementary Acknowledgements:**

We wish to acknowledge all of the investigators and funding agencies that enabled the deposition of data in EGA and dbGaP that we used in this study:

This study makes use data generated by the Wellcome Trust Case-Control Consortium. A full list of the investigators who contributed to the generation of the data is available from www.wtccc.org.uk. Funding for the WTCCC project was provided by the Wellcome Trust under award 076113 and 085475.

Funding support for the GWAS of Gene and Environment Initiatives in Type 2 Diabetes was provided through the NIH Genes, Environment and Health Initiative [GEI] (U01HG004399). The human subjects participating in the GWAS derive from The Nurses' Health Study and Health Professionals' Follow-up Study and these studies are supported by National Institutes of Health grants CA87969, CA55075, and DK58845. Assistance with phenotype harmonization and genotype cleaning, as well as with general study coordination, was provided by the Gene Environment Association Studies, GENEVA Coordinating Center (U01 HG004446). Assistance with data cleaning was provided by the National Center for Biotechnology Information. Funding support for genotyping, which was performed at the Broad Institute of MIT and Harvard, was provided by the NIH GEI (U01HG004424). The datasets used for the analyses described in this manuscript were obtained from dbGaP at [http://www.ncbi.nlm.nih.gov/projects/gap/cgi-bin/study.cgi?study_id=phs000091.v1.p1] through dbGaP accession number [phs000091.v2.p1].

This work utilized in part data from the NINDS DbGaP database from the CIDR:NGRC PARKINSON'S DISEASE STUDY through dbGaP accession number phs000196.v2.p1.

Funding support for NINDS Parkinsons Disease was provided by the National Institute for Neurological Disease and Stroke (NINDS) and the genotyping of samples was provided by the Singleton Lab (NIA Laboratory of Neurogenetics)with support from NINDS. The dataset(s) used for the analyses described in this manuscript were obtained from the NINDS Database found at http://view.ncbi.nlm.nih.gov/dbgap through dbGaP accession number phs000089.v3.p2.

Funding support for the companion studies, Genome-Wide Association Study of Schizophrenia (GAIN) and Molecular Genetics of Schizophrenia - nonGAIN Sample (MGS_nonGAIN), was provided by Genomics Research Branch at NIMH see below) and the genotyping and analysis of samples was provided through the Genetic Association Information Network (GAIN) and under the MGS U01s: MH79469 and MH79470. Assistance with data cleaning was provided by the National Center for Biotechnology Information. The MGS dataset(s) used for the analyses described in this manuscript were obtained from the database of Genotype and Phenotype (dbGaP) found at http://www.ncbi.nlm.nih.gov/gap through dbGaP accession numbers phs000021.v2.p1 (GAIN) and phs000167.v1.p1 (nonGAIN). Samples and associated phenotype data for the MGS GWAS study were collected under the following grants: NIMH Schizophrenia Genetics Initiative U01s: MH46276 (CR Cloninger), MH46289 (C Kaufmann), and MH46318 (MT Tsuang); and MGS Part 1 (MGS1) and Part 2 (MGS2) R01s: MH67257 (NG Buccola), MH59588 (BJ Mowry), MH59571 (PV Gejman), MH59565 (Robert Freedman), MH59587 (F Amin), MH60870 (WF Byerley), MH59566 (DW Black), MH59586 (JM Silverman), MH61675 (DF Levinson), and MH60879 (CR Cloninger). Further details of collection sites, individuals, and institutions may be found in data supplement Table 1 of Sanders et al. (2008; PMID: 18198266) and at the study dbGaP pages."

Funding support for the GAIN Major Depression: Stage 1 Genome-wide Association In Population Based Samples Study (parent studies: Netherlands Study of Depression and Anxiety (NESDA) and the Netherlands Twin Register (NTR)) was provided by the Netherlands Scientific Organization (904-61-090, 904-61-193, 480-04-004, 400-05-717, NWO Genomics, SPI 56-464- 1419) the Centre for Neurogenomics and Cognitive Research (CNCR-VU); the European Union (EU/WLRT-2001-01254), ZonMW (geestkracht program, 10-000-1002), NIMH (RO1 MH059160) and matching funds from participating institutes in NESDA and NTR, and the genotyping of samples was provided through the Genetic Association Information Network (GAIN). The dataset(s) used for the analyses described in this manuscript were obtained from the database of Genotypes and Phenotypes (dbGaP) found at http://www.ncbi.nlm.nih.gov/gap through dbGaP accession number phs000020.v2.p1. Samples and associated phenotype data for the GAIN Major Depression: Stage 1 Genome-wide Association In Population Based Samples Study (PI: Dr. Patrick F. Sullivan, MD, University of North Carolina) were provided by Dr. Dorret I. Boomsma, PhD and Dr. Eco de Geus, PhD VU University Amsterdam (PIs NTR), Dr. Brenda W. Penninx, PhD, VU University Medical Center Amsterdam, Dr. Frans Zitman, MD PhD, Leiden University Medical Center, Leiden, and Dr. Willem Nolen, MD PhD, University Medical Center Groningen (PIs and site-PIs NESDA).

Funding support for the Whole Genome Association Study of Bipolar Disorder was provided by the National Institute of Mental Health (NIMH) and the genotyping of samples was provided through the Genetic Association Information Network (GAIN). The datasets used for the analyses described in this manuscript were obtained from the database of Genotypes and Phenotypes (dbGaP) found at http://www.ncbi.nlm.nih.gov/gap through dbGaP accession number phs000017.v3.p1. Samples and associated phenotype data for the Collaborative Genomic Study of Bipolar Disorder were provided by the The NIMH Genetics Initiative for Bipolar Disorder. Data and biomaterials were collected in four projects that participated in NIMH Bipolar Disorder Genetics Initiative. From 1991-98, the Principal Investigators and Co-Investigators were: Indiana University, Indianapolis, IN, U01 MH46282, John Nurnberger, M.D., Ph.D., Marvin Miller, M.D., and Elizabeth Bowman, M.D.; Washington University, St. Louis, MO, U01 MH46280, Theodore Reich, M.D., Allison Goate, Ph.D., and John Rice, Ph.D.; Johns Hopkins University, Baltimore, MD U01 MH46274, J. Raymond DePaulo, Jr., M.D., Sylvia Simpson, M.D., MPH, and Colin Stine, Ph.D.; NIMH Intramural Research Program, Clinical Neurogenetics Branch, Bethesda, MD, Elliot Gershon, M.D., Diane Kazuba, B.A., and Elizabeth Maxwell, M.S.W. Data and biomaterials were collected as part of ten projects that participated in the NIMH Bipolar Disorder Genetics Initiative. From 1999-03, the Principal Investigators and Co-Investigators were: Indiana University, Indianapolis, IN, R01 MH59545, John Nurnberger, M.D., Ph.D., Marvin J. Miller, M.D., Elizabeth S. Bowman, M.D., N. Leela Rau, M.D., P. Ryan Moe, M.D., Nalini Samavedy, M.D., Rif El-Mallakh, M.D. (at University of Louisville), Husseini Manji, M.D. (at Wayne State University), Debra A. Glitz, M.D. (at Wayne State University), Eric T. Meyer, M.S., Carrie Smiley, R.N., Tatiana Foroud, Ph.D., Leah Flury, M.S., Danielle M. Dick, Ph.D., Howard Edenberg, Ph.D.; Washington University, St. Louis, MO, R01 MH059534, John Rice, Ph.D, Theodore Reich, M.D., Allison Goate, Ph.D., Laura Bierut, M.D. ; Johns Hopkins University, Baltimore, MD, R01 MH59533, Melvin McInnis M.D. , J. Raymond DePaulo, Jr., M.D., Dean F. MacKinnon, M.D., Francis M. Mondimore, M.D., James B. Potash, M.D., Peter P. Zandi, Ph.D, Dimitrios Avramopoulos, and Jennifer Payne; University of Pennsylvania, PA, R01 MH59553, Wade Berrettini M.D.,Ph.D. ; University of California at Irvine, CA, R01 MH60068, William Byerley M.D., and Mark Vawter M.D. ; University of Iowa, IA, R01 MH059548, William Coryell M.D. , and Raymond Crowe M.D. ; University of Chicago, IL, R01 MH59535, Elliot Gershon, M.D., Judith Badner Ph.D. , Francis McMahon M.D. , Chunyu Liu Ph.D., Alan Sanders M.D., Maria Caserta, Steven Dinwiddie M.D., Tu Nguyen, Donna Harakal; University of California at San Diego, CA, R01 MH59567, John Kelsoe, M.D., Rebecca McKinney, B.A.; Rush University, IL, R01 MH059556, William Scheftner M.D. , Howard M. Kravitz, D.O., M.P.H., Diana Marta, B.S., Annette Vaughn-Brown, MSN, RN, and Laurie Bederow, MA; NIMH

Intramural Research Program, Bethesda, MD, 1Z01MH002810-01, Francis J. McMahon, M.D., Layla Kassem, PsyD, Sevilla Detera-Wadleigh, Ph.D, Lisa Austin,Ph.D, Dennis L. Murphy, M.D.

Funding support for the GWAS of Lung Cancer and Smoking was provided through the NIH Genes, Environment and Health Initiative [GEI] (Z01 CP 010200). The human subjects participating in the GWAS derive from The Environment and Genetics in Lung Cancer Etiology (EAGLE) case-control study and the Prostate, Lung Colon and Ovary Screening Trial and these studies are supported by intramural resources of the National Cancer Institute. Assistance with phenotype harmonization and genotype cleaning, as well as with general study coordination, was provided by the Gene Environment Association Studies, GENEVA Coordinating Center (U01 HG004446). Assistance with data cleaning was provided by the National Center for Biotechnology Information. Funding support for genotyping, which was performed at the Johns Hopkins University Center for Inherited Disease Research, was provided by the NIH GEI (U01HG004438). The datasets used for the analyses described in this manuscript were obtained from dbGaP at http://www.ncbi.nlm.nih.gov/gap through dbGaP accession number phs000336.

The CGEMS Breast Cancer study was supported by the National Cancer Institute. Details of the study can be found at http://cgems.cancer.gov/data and in reference 28. The data was obtained from dbGaP under accession phs000147v1.

Funding support for the Study of Addiction: Genetics and Environment (SAGE) was provided through the NIH Genes, Environment and Health Initiative [GEI] (U01 HG004422). SAGE is one of the genome-wide association studies funded as part of the Gene Environment Association Studies (GENEVA) under GEI. Assistance with phenotype harmonization and genotype cleaning, as well as with general study coordination, was provided by the GENEVA Coordinating Center (U01 HG004446). Assistance with data cleaning was provided by the National Center for Biotechnology Information. Support for collection of datasets and samples was provided by the Collaborative Study on the Genetics of Alcoholism (COGA; U10 AA008401), the Collaborative Genetic Study of Nicotine Dependence (COGEND; P01 CA089392), and the Family Study of Cocaine Dependence (FSCD; R01 DA013423). Funding support for genotyping, which was performed at the Johns Hopkins University Center for Inherited Disease Research, was provided by the NIH GEI (U01HG004438), the National Institute on Alcohol Abuse and Alcoholism, the National Institute on Drug Abuse, and the NIH contract "High throughput genotyping for studying the genetic contributions to human disease" (HHSN268200782096C). The datasets used for the analyses described in this manuscript were obtained from dbGaP at http://www.ncbi.nlm.nih.gov/projects/gap/cgi- bin/study.cgi?study_id=phs000092.v1.p1 through dbGaP accession number phs000092.v1.p1.

Supplementary Figure S1. MAC distribution profiles among the strains in a RILs or segregants panel

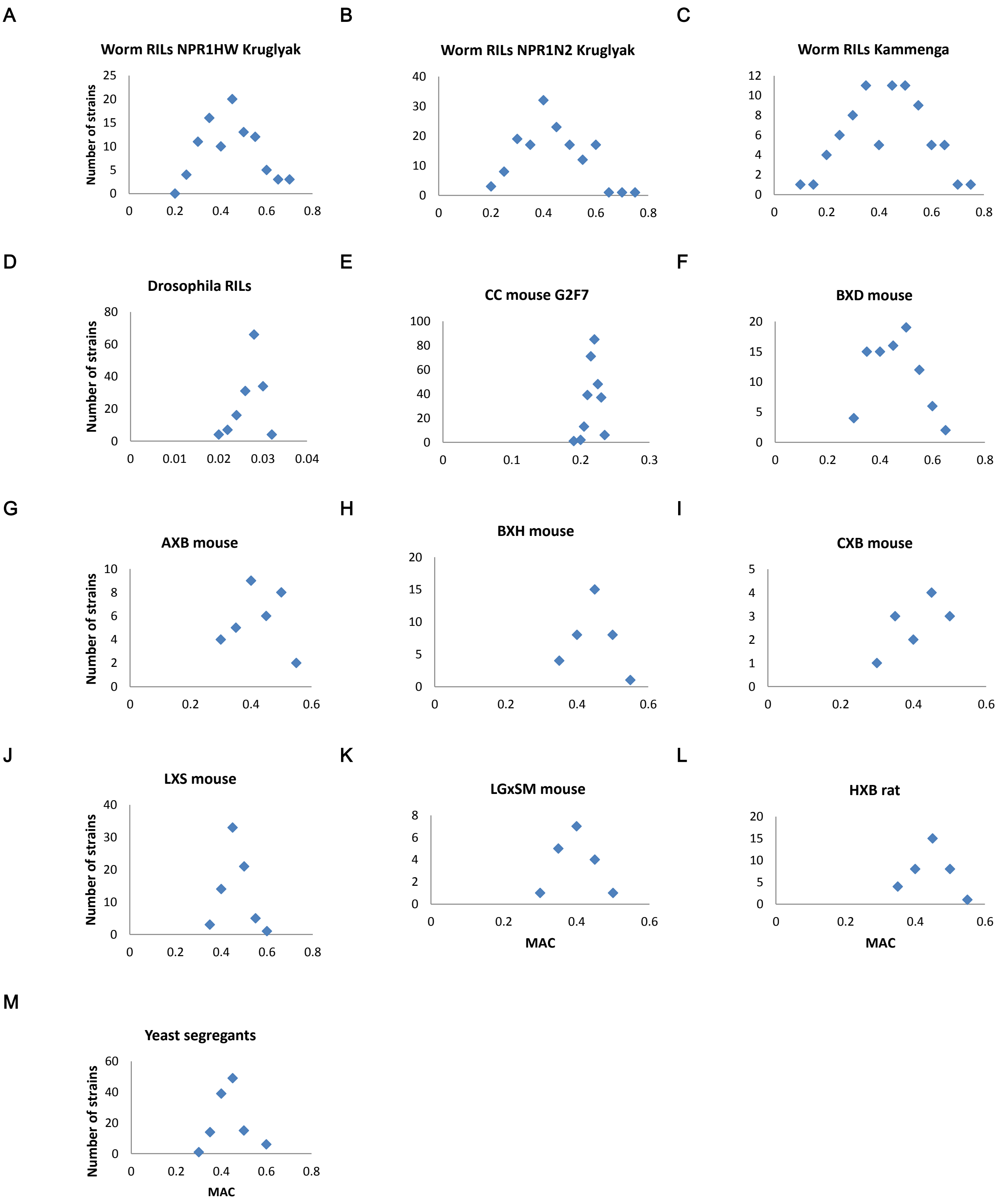

Supplementary Figure S2.

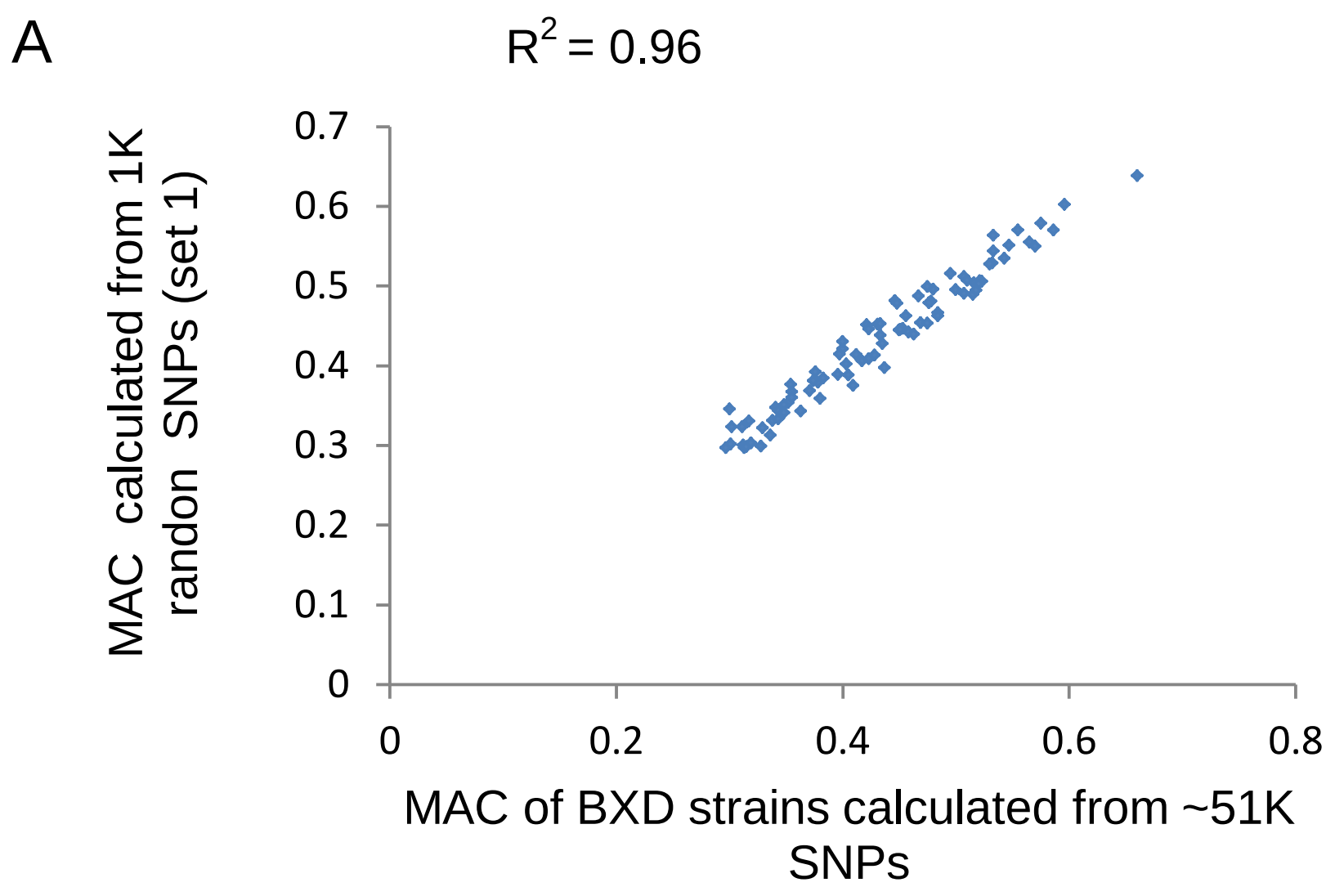

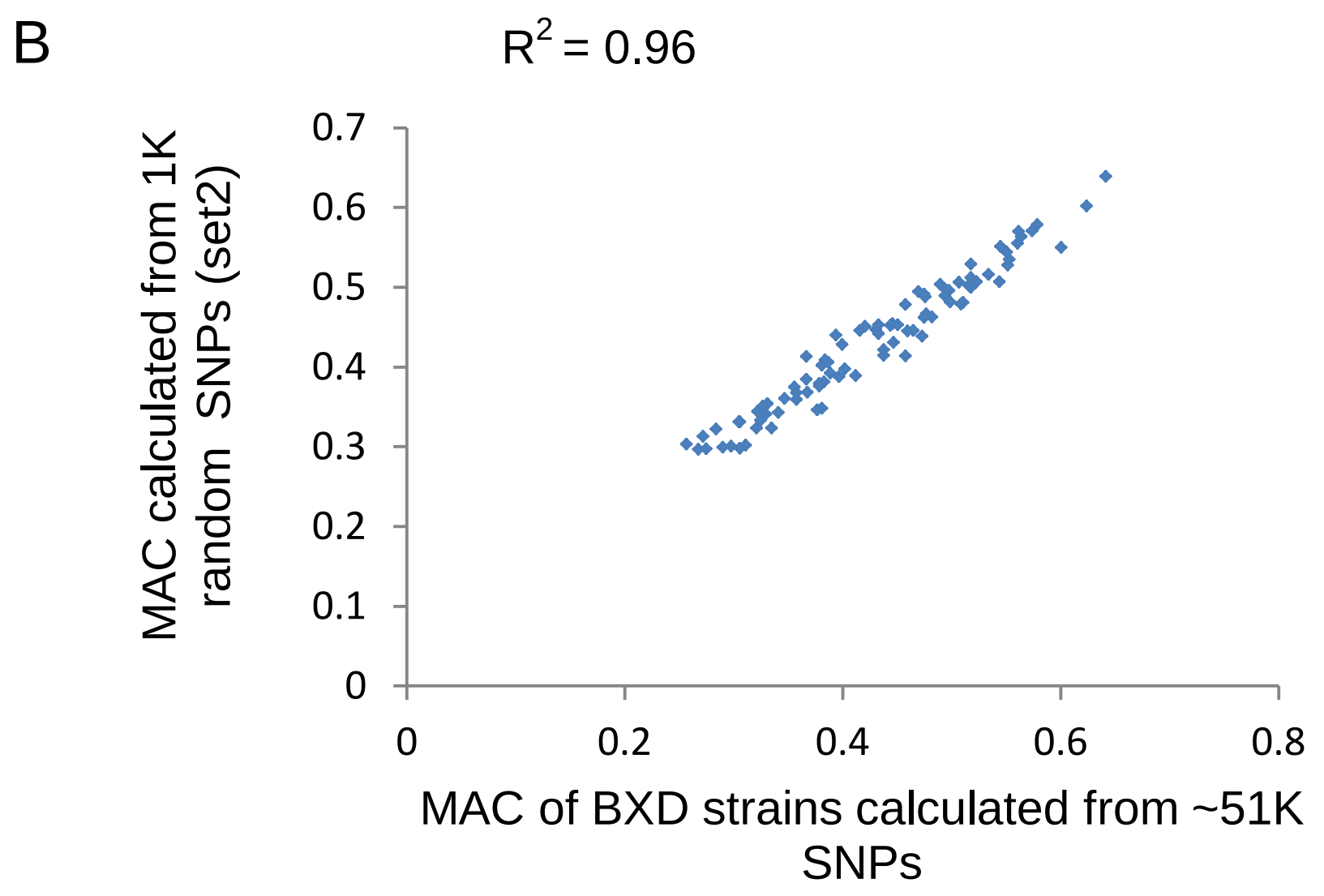

**Supplementary Table S2. Selected traits consistently linked (or not) with higher MAC in different panels of RILs**

| | | Reprod. fit.[a] | | | Life span | | | Alcohol sens.[b] | | | Cancer[c] | | | Blood pressure. | | |
|---|---|---|---|---|---|---|---|---|---|---|---|---|---|---|---|---|
| RIL panel | Correlation | r | *p* | # | r | *p* | # | r | *p* | # | r | *p* | # | r | *p* | # |
| BXD | Pearson | -0.23 | ns | 61 | -0.12 | ns | 20 | -0.27 | 0.04 | 25 | 0.22 | ns | 22 | -0.27 | ns | 21 |
| | Spearman | -0.25 | 0.05 | | -0.16 | ns | | -0.25 | 0.05 | | 0.13 | ns | | -0.28 | ns | |
| AXBBXA | Pearson | | | | | | | | | | t test, p<0.05 | | 17 | | | |
| | Spearman | | | | | | | | | | Lung tumor | | | | | |
| CXB | Pearson | | | | | | | | | | 0.69 | ns | 7 | | | |
| | Spearman | | | | | | | | | | 0.82 | 0.03 | | | | |
| LXS | Pearson | | | | -0.25 | ns | 43 | -0.23 | 0.05 | 74 | | | | | | |
| | Spearman | | | | -0.29 | 0.06 | | -0.27 | 0.02 | | | | | | | |
| CC (F6) | Pearson | -0.14 | 0.04 | 245 | | | | | | | | | | | | |
| | Spearman | -0.13 | ns | | | | | | | | | | | | | |
| BXHHXB | Pearson | -0.49 | 0.02 | 24 | | | | | | | | | | -0.02 | ns | 32 |
| | Spearman | -0.50 | 0.01 | | | | | | | | | | | -0.06 | ns | |
| Worm | Pearson | -0.45 | 0.002 | 42 | | | | | | | | | | | | |
| Kruglyak | Spearman | -0.45 | 0.002 | | | | | | | | | | | | | |
| Worm | Pearson | | | | -0.27 | ns | 35 | | | | | | | | | |
| Kammenga | Spearman | | | | -0.35 | 0.04 | | | | | | | | | | |

[a]Reproductive fitness includes uterus horn length (BXD), litter/brood size (CC, Worm), and fetal weights in left horn of uterus (BXHHXB rat).

[b]Alcohol sensitity was assayed by distance traveled after ethanol injection.

[c]Cancer includes DEN induced liver tumor (BXD), urethane induced lung tumor (AXBBXA), and virus induced lymphoma (CXB).

**Supplementary Table S4. Correlations among selected traits linked with higher MAC in BXD mice**

| | Sample | Pearson | | | | | | | | | | | | | | | | | | | | | | |
|---|---|---|---|---|---|---|---|---|---|---|---|---|---|---|---|---|---|---|---|---|---|---|---|---|
| GN ID[a] | # | r | *p* | Trait | 1 | 2 | 3 | 4 | 5 | 6 | 7 | 8 | 9 | 10 | 11 | 12 | 13 | 14 | 15 | 16 | 17 | 18 | 19 | Trait description |
| 10145 | 25 | -0.63 | 0.001 | 1 | | | | | | ** | | *[b] | | | | | | | | | | | * | Maxi-threshold to ethanol induced ataxia |
| 10169 | 25 | -0.63 | 0.001 | 2 | | | * | * | | | | * | | | | | | | | | | | | Methamphet. induced temp. change |
| 11453 | 61 | -0.42 | 0.001 | 3 | | | | | | | | * | | * | *** | | | | | ** | | | | Blood ethanol concentration for males |
| 11307 | 60 | -0.41 | 0.001 | 4 | | | | | ** | | | | | | | | ** | * | * | | | | | Hargreaves' test for males |
| 11672 | 57 | -0.42 | 0.001 | 5 | | | | | | | | | | | | | | | * | | | | * | Open field rearing activity from 10-15 min |
| 10022 | 20 | 0.60 | 0.006 | 6 | | | | | | | ** | | ** | | | | | | | | ** | | | Saccharin preference versus water ratio |
| 10493 | 25 | -0.47 | 0.017 | 7 | | | | | | | | | * | | | | | | | | | | | Cocaine induced difference in locomotion |
| 10301 | 23 | 0.53 | 0.010 | 8 | | | | | | | | | | | | | | | * | | | * | | Cocaine, nose pokes in hole board |
| 10494 | 24 | -0.51 | 0.012 | 9 | | | | | | | | | | | | | | | | * | | | | Ethanol induced difference in locomotion |
| 10917 | 15 | -0.68 | 0.006 | 10 | | | | | | | | | | | | ** | | | | | | | | Anxiety, transitions between light and dark |
| 14220 | 28 | 0.51 | 0.005 | 11 | | | | | | | | | | | | | * | | | ** | | * | | Resistin level after high fat diet |
| 12540 | 22 | -0.69 | 0.0004 | 12 | | | | | | | | | | | | | | | | | | | | Transferrin saturation fed 3 ppm iron diet |
| 11725 | 57 | 0.50 | 0.007 | 13 | | | | | | | | | | | | | | | | | | | * | Gain in weight between 8 and 9 wks |
| 12886 | 16 | 0.64 | 0.008 | 14 | | | | | | | | | | | | | | | * | | | | | Oxygen consumption males |
| 12568 | 38 | 0.57 | 0.0001 | 15 | | | | | | | | | | | | | | | | | | ** | | Deoxycorticosterone in cerebral cortex |
| 12852 | 16 | 0.68 | 0.004 | 16 | | | | | | | | | | | | | | | | | | * | | Food intake of 13-week old females |
| 12554 | 24 | 0.38 | 0.070 | 17 | | | | | | | | | | | | | | | | | | | | Depression assay, duration of immobility |
| 14226 | 28 | -0.45 | 0.017 | 18 | | | | | | | | | | | | | | | | | | | | IL17 level after high fat diet |
| 12396 | 64 | 0.28 | 0.026 | 19 | | | | | | | | | | | | | | | | | | | | Time in open quadr. in elevated 0 maze |

[a]GeneNetwork identification number.

[b]Symbols *, **, and *** represent $p < 0.05$, 0.01, 0.001 respectively, from Pearson analysis.

**Supplementary Table S6. MAC correlation with yeast growth in the presence of a compound using 105 segregants.**

| | Pearson | | Spearman | |
|---|---|---|---|---|
| Compounds | r | *p* | r | *p* |
| diphenyleneiodonium 64 h 16 µM | -0.44 | <0.0001 | -0.39 | <0.0001 |
| *sertraline 68 h 20.9 µM*[a] | -0.36 | 0.0004 | -0.36 | 0.0003 |
| *sertraline 52 h 20.9 µM* | -0.30 | 0.003 | -0.30 | 0.004 |
| *sertraline 78 h 20.9 µM* | -0.31 | 0.002 | -0.33 | 0.001 |
| diphenyleneiodonium 64 h 16 µM | -0.30 | 0.003 | -0.29 | 0.004 |
| alverine 118 h 105.5 µM | -0.29 | 0.004 | -0.31 | 0.002 |
| *tamoxifen 70 h 13.5 µM* | -0.33 | 0.001 | -0.28 | 0.007 |
| *trimeprazine 80 h 83.8 µM* | -0.33 | 0.001 | -0.33 | 0.001 |
| *chlorpromazine 70 h 15.7 µM* | -0.31 | 0.003 | -0.27 | 0.009 |
| *sertraline 90 h 20.9 µM* | -0.31 | 0.002 | -0.32 | 0.002 |
| cinnarazine 68 h 33.9 µM | -0.27 | 0.009 | -0.22 | 0.03 |
| *trifluoperazine 90 h 26 µM* | -0.27 | 0.007 | -0.27 | 0.007 |

[a]FDA-approved drugs in italics.

**Supplementary Table S7. Correlations between MAC and clinical traits.**

| | Pearson | | Spearman | |
|---|---|---|---|---|
| | r | *p* | r | *p* |
| Alcohol addiction cases (1167 sub. Ave. age 38.1 yr) | | | | |
| Cocaine dependence | 0.11 | 0.0002 | 0.08 | 0.008 |
| Cocaine_sx1 (tolerance) | 0.09 | 0.001 | 0.06 | 0.04 |
| Cocaine_sx2 (withdrawal) | 0.09 | 0.002 | 0.07 | 0.02 |
| Cocaine_sx3 (more than intended) | 0.09 | 0.002 | 0.07 | 0.01 |
| Cocaine_sx4 (desire to cut) | 0.11 | 0.0003 | 0.07 | 0.02 |
| Cocaine_sx5 (time spent to use) | 0.11 | 0.0002 | 0.08 | 0.005 |
| Cocaine_sx6 (reduced social act.) | 0.08 | 0.009 | 0.07 | 0.03 |
| Sex (female 2, male 1) | -0.13 | < 0.0001 | -0.11 | 0.0001 |
| Trauma sexual | -0.09 | 0.004 | -0.07 | 0.04 |
| Education level | -0.08 | 0.004 | -0.06 | 0.03 |
| Cigarette daily | -0.08 | -0.003 | 0.06 | 0.06 |
| Alcohol addiction controls (1366 sub. Ave. age 38.6 yr) | | | | |
| Education level | -0.12 | < 0.0001 | -0.05 | 0.08 |
| Income | -0.13 | < 0.0001 | -0.09 | 0.002 |
| Height | -0.12 | < 0.0001 | -0.09 | 0.002 |
| Nic_sx3 (use more than intended) | -0.07 | 0.02 | -0.05 | 0.05 |
| Alc_sx3 (use more than intended) | -0.10 | 0.0001 | -0.09 | 0.0005 |
| Cigarette daily | -0.05 | 0.05 | -0.05 | 0.07 |
| PD controls (ave. age 72.8 yr) | | | | |
| Life span cohort 1 (46 sub, ave. age 88.2 yr) | -0.30 | <0.05 | ns | |
| Life span cohort 2 (34 sub, ave. age 87.2 yr) | ns | | -0.40 | <0.05 |
| NSAIDs OTC load, cohort 1 (999 sub.) | 0.08 | 0.1 (ns) | ns | |
| NSAIDs OTC load, cohort 2 (993 sub.) | 0.14 | 0.003 | ns | |